\documentclass[useAMS,usenatbib]{mn2e}
\usepackage{graphicx,amssymb, txfonts,pstricks}

\def\pb{Pa$\beta$}
\def\br{Br$\gamma$}
\def\ha{H$\alpha$}
\def\hb{H$\beta$}
\def\feii{[Fe\,{\sc ii}]}
\def\pii{[P\,{\sc ii}]}
\def\oiii{[O\,{\sc iii}]}
\def\oi{[O\,{\sc i}]}
\def\sii{[S\,{\sc ii}]}
\def\nii{[N\,{\sc ii}]}
\def\caviii{[Ca\,{\sc viii}]}
\def\h2{H$_2$}
\def\p1{Paper~I}
\def\kl{K$_{\rm l}$}
\def\kms {$\rm km\,s^{-1}$}

\title[Compact molecular disc in Mrk~1066]{Compact molecular disc and ionized gas outflows within 350~pc of the active nucleus of  Mrk~1066}
\author[Riffel and Storchi-Bergmann]{Rogemar. A. Riffel$^{1}$\thanks{E-mail:
rogemar@smail.ufsm.br (RAR); thaisal@ufrgs.br (TSB)} and Thaisa Storchi-Bergmann
$^{2}$\\
$^{1}$ Universidade Federal de Santa Maria, Departamento de F\'\i sica, Centro de Ci\^encias Naturais e Exatas, 97105-900, Santa Maria, RS, Brazil \\ 
$^{2}$Universidade Federal do Rio Grande do Sul, Instituto de F\'\i sica, CP 15051, Porto Alegre 91501-970, RS, Brazil
}
\begin{document}

\date{Accepted 1988 December 15. Received 1988 December 14; in original form 1988 October 11}

\pagerange{\pageref{firstpage}--\pageref{lastpage}} \pubyear{2002}

\maketitle

\label{firstpage}

\begin{abstract}

We present stellar and gaseous kinematics of the inner $\approx$\,350~pc radius of the Seyfert galaxy Mrk\,1066 derived from J and K$_l$ bands data obtained with the Gemini's Near-Infrared Integral Field Spectrograph (NIFS) at a spatial resolution of  $\approx$35\,pc. The stellar velocity field is dominated by rotation in the galaxy plane but shows an S-shape distortion along the galaxy minor axis which seems to be due to an oval structure seen in an optical continuum image. Along this oval, between 170 and 280~pc from the nucleus we find a partial ring of low $\sigma_*$ ($\approx$50\,\kms) attributed to an intermediate age stellar population. The velocity dispersion of the stellar bulge ($\sigma_*\,\approx90$\,\kms) implies a super-massive black hole mass  of $\approx 5.4 \times 10^6$\,M$_\odot$.  From measurements of the emission-line fluxes and profiles (\pii$\lambda1.1886\,\mu$m, \feii$\lambda1.2570\,\mu$m, \pb\ and \h2$\lambda2.1218\,\mu$m),  we have constructed maps for the gas centroid velocity, velocity dispersion, as well as channel maps. The velocity fields for all emission lines are dominated by a similar rotation pattern to that observed for the stars, but are distorted by the presence of two structures:  (i) a compact rotating disc with radius $r\approx$\,70\,pc; (ii) outflows along the radio jet which is oriented approximately along the galaxy major axis. The compact rotating disc is more conspicuous in the \h2\ emitting gas, which presents the smallest $\sigma$ values ($\le70$\,\kms) and most clear rotation pattern, supporting a location in the galaxy plane. We estimate a gas mass for the disc of  $\sim\,10^7\,{\rm M_\odot}$. The \h2\ kinematics further suggests that the nuclear disc is being fed by gas coming from the outer regions.  The outflow is more conspicuous in the \feii\ emitting gas, which presents the highest $\sigma$ values (up to 150\,\kms) and the highest blue and redshifts of up to 500\,\kms, while the highest stellar rotation velocity is only $\approx$\,130\,\kms. We estimate a mass-outflow rate in ionized gas of $\approx6\times10^{-2}{\rm M_\odot yr^{-1}}$. The derived kinematics for the emitting gas is similar to that observed in previous studies supporting that the \h2\ is a tracer of the AGN feeding and the \feii\  of its feedback.

\end{abstract}

\begin{keywords}
galaxies: individual (Mrk\,1066) -- galaxies: Seyfert -- galaxies: ISM -- galaxies: kinematics and dynamics --
 infrared: galaxies
\end{keywords}

\section{Introduction}

This work is a continuing study of the central region of Mrk\,1066 using Gemini's Near-infrared Integral Field Spectrograph (NIFS) observations in the J and K$_{\rm l}$ bands. In \citet[][hereafter \p1]{paper1}, we presented maps for the emission-line flux distribution and ratios and discussed the gas excitation and extinction, as well as the origin of the nuclear continuum.  The main results of \p1\ are: (i) The line emission in the near-IR lines is extended over the whole NIFS field ($\approx$\,700\,pc) and most elongated at position angle PA=135/315$^\circ$, showing 
a good correlation with the optical \oiii\ line and radio continuum emission. [Ca\,{\sc viii}]$\,\lambda2.3220\,\mu$m  and [S\,{\sc ix}]$\,\lambda1.2524\,\mu$m coronal lines 
are the exceptions, being unresolved at the nucleus;  
(ii)  The \feii$\,\lambda1.2570\,\mu$m/\pb~$vs$~\h2$\,\lambda2.1218\,\mu$m/\br\ diagnostic diagram is dominated by values typical of active galaxies; (iii)  The reddening map obtained from the \pb/\br~line ratio presents a S-shape structure with $E(B-V)$ reaching a value of 1.7 along PA$\approx$135/315$^\circ$;
(iv) From line ratio maps we conclude that the main excitation mechanism of the \h2\ and \feii\ emission lines  
is heating by X-rays from the central AGN. Correlations between the radio and line-emission maps  -- stronger for  \feii\ than for the \h2\ -- suggest that shocks due to the radio jet play a role in the gas excitation; (v) The nucleus contains an unresolved infrared source whose continuum is well reproduced by emission from dust with temperature 
$\sim$810\,K and mass $\sim1.4\times10^{-2}\,{\rm M_\odot}$. In \citet{paper2} we presented stellar population synthesis, which shows that within $\sim$160\,pc from the nucleus an old stellar population dominates the near-IR continuum (age $\sim$10$^{10}$\,yr), while in a partial ring surrounding this region, the continuum is dominated by an intermediate age stellar population ($10^8\le{\rm age}\le\,10^9$\,yr).

Mrk\,1066 is an SB0 galaxy located at a distance of 48.6\,Mpc which harbours a Seyfert 2 nucleus\footnote{NASA/IPAC Extragalactic Database (NED -- 
http://nedwww.ipac.caltech.edu)}, for which 1$^{\prime\prime}$ 
corresponds to 235\,pc at the galaxy. Previous optical long-slit spectroscopy shows that the high-excitation gas (traced by the \oiii\ emission) and 
low-excitation gas (traced by \nii, \sii, \oi\ and H recombination lines) present distinct kinematics, with the former being more disturbed with respect to the rotation curve observed in the low-excitation emission lines \citep{bower95}. The authors suggest that the high-excitation gas kinematics is due to an outflow driven by the
radio jet, which is co-spatial with the extended line emission, along position angle PA=135/315$^\circ$ \citep{ulvestad89,nagar99}. 
A distinct kinematics for the high- and low- excitation gas is also supported by optical 
integral field spectroscopy of the inner 1~kpc of Mrk\,1066 \citep{stoklasova09}. In the near-IR,
the emission lines present asymmetric profiles revealed by long-slit observations \citep{knop01}. 
These  asymmetries have been interpreted has being originated in the same outflowing gas component observed in \oiii, while most of the near-IR line 
emission (traced by the peak of the line profiles) may be due to emission from gas located in the galaxy plane \citep{knop01}. 

In the present paper, we use the observations described in \p1 to obtain two-dimensional maps for the gaseous and stellar kinematics of the inner $\approx\,350\,$pc radius of Mrk\,1066, which allowed us to put additional constraints on the physical scenario for the circumnuclear region 
of this galaxy. This paper is organized as follows: In Section~2 we present the description of the observations and data reduction,
 Sec.~3 presents the results for the gaseous and stellar kinematics, which are discussed in Sec.~4. Sec.~5 presents the conclusions of this work.

\begin{figure}  
 \centering
 \begin{minipage}{1\linewidth}
 \includegraphics[scale=0.22]{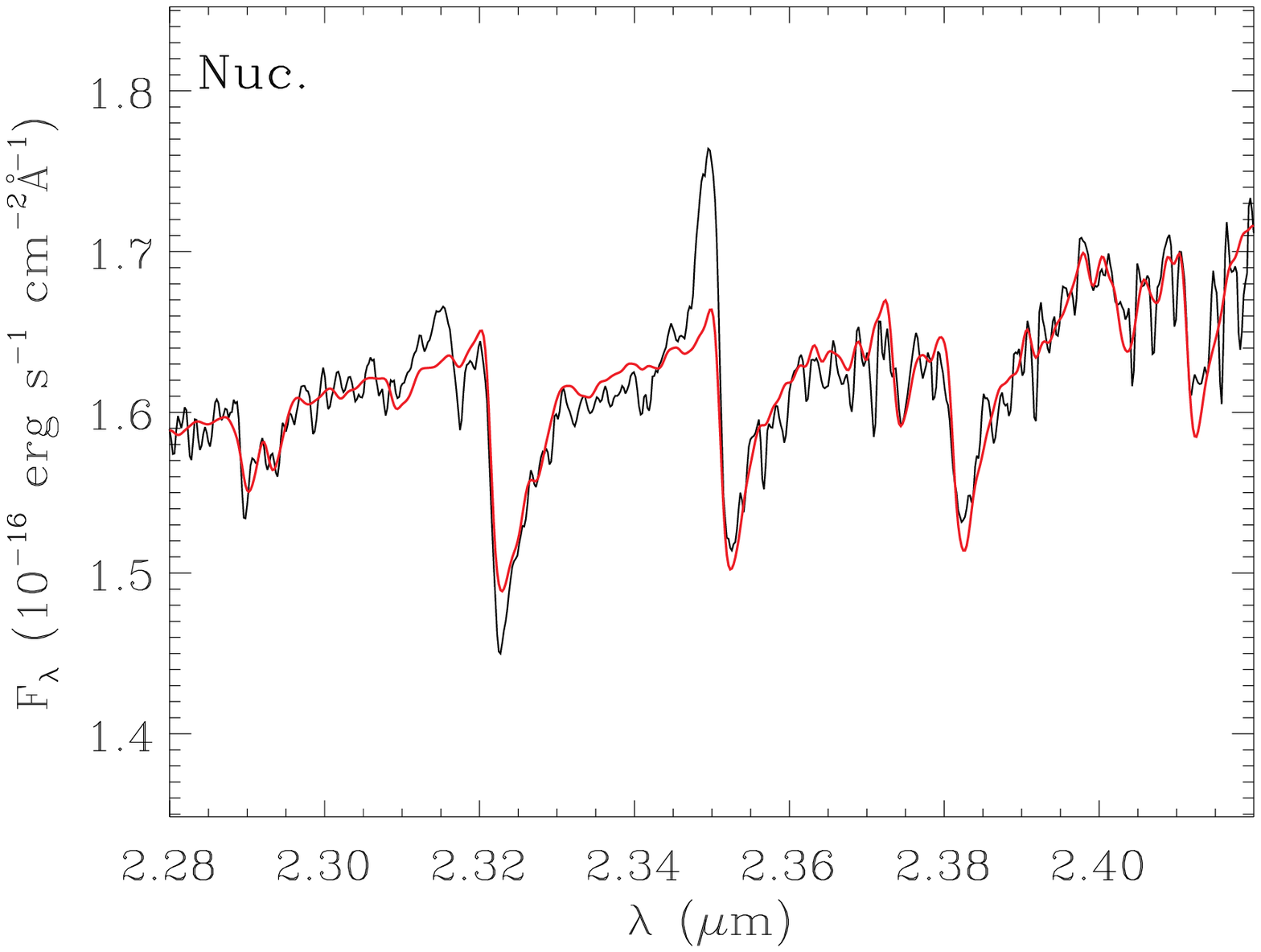}
 \includegraphics[scale=0.22]{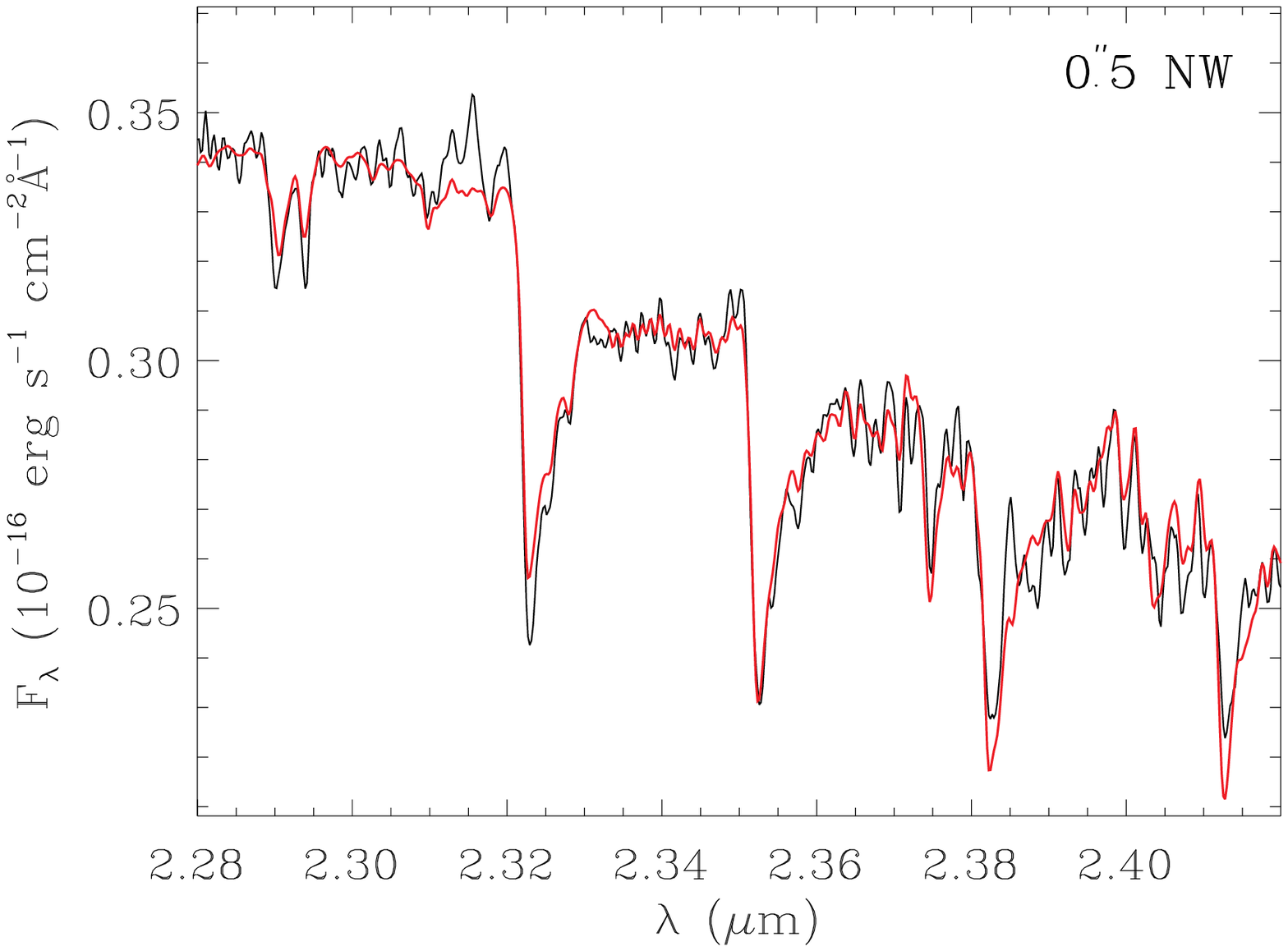}
 \end{minipage}
 \begin{minipage}{1\linewidth}
 \includegraphics[scale=0.22]{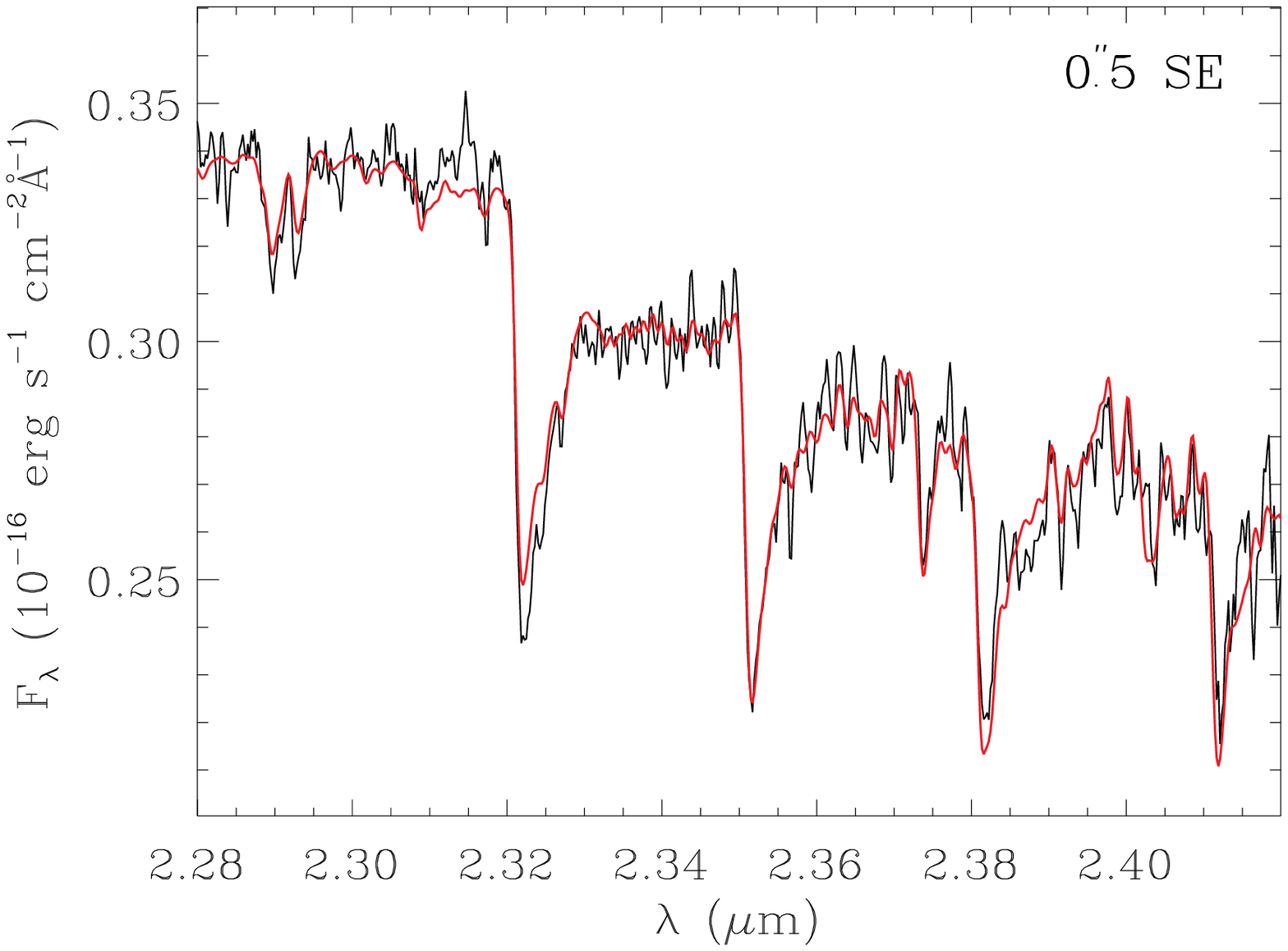}
 \includegraphics[scale=0.22]{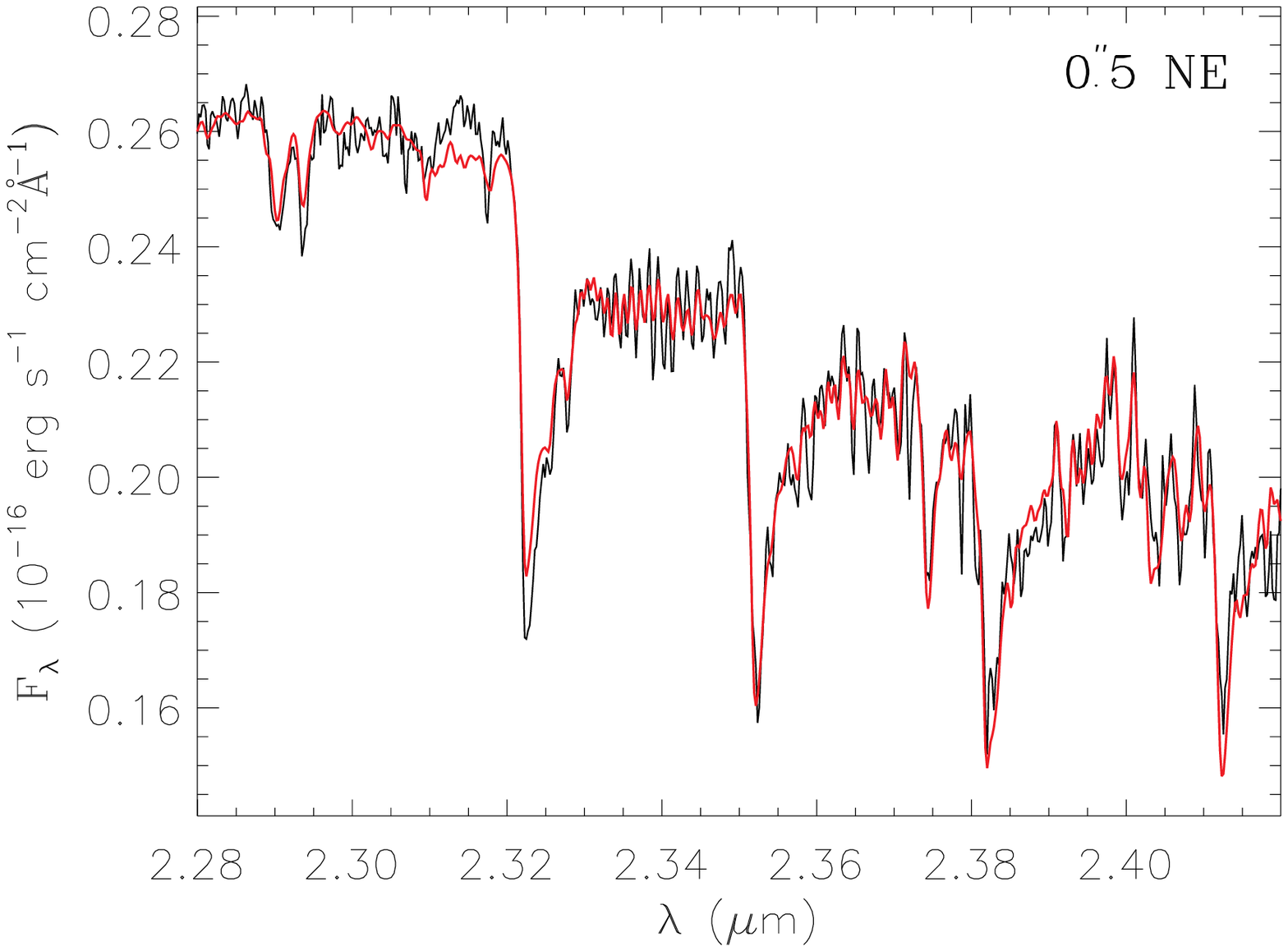}
 \end{minipage}
 \caption{Sample fits of the stellar kinematics of the nuclear region of Mrk\,1066 using pPXF. Top left-hand panel: fit of the nuclear spectrum; top right-hand panel:
fit of a spectrum at 0\farcs5 north-west from the nucleus;  bottom left-hand panel: 
fit of a spectrum at 0\farcs5 south-east from the nucleus;  bottom right-hand panel:
fit of a spectrum at 0\farcs5 north-east from the nucleus. The observed spectra is shown in black and the resulting fit in red.} 
 \label{stel_fit}  
 \end{figure}

\section{Observations and Data Reduction}

The J and K$_{\rm l}$ bands Integral Field Unit (IFU) spectroscopic data of Mrk\,1066 were obtained with Gemini NIFS \citep{mcgregor03} 
operating with the Gemini North Adaptive Optics system ALTAIR in September 2008 under the programme GN-2008B-Q-30. NIFS has a 
square field of view of $\approx3\farcs0\times3\farcs0$, divided into 29 slices with an 
angular sampling of 0$\farcs$103$\times$0$\farcs$042, optimized for use with ALTAIR. 

The J band observations were centred at 1.25\,$\mu$m, covering a spectral region from 1.15\,$\mu$m to 1.36\,$\mu$m at a spectral resolution 
of  $\approx1.7\,\AA$ (from the full width at half maximum -- FWHM -- of arc lamp lines) and at spatial resolution of  0\farcs13$\pm$0\farcs02, as obtained from 
the FWHM of the spatial profile of the telluric star. The \kl~observations, centred at 2.3\,$\mu$m  covered the spectral range from  2.10$\,\mu$m to 2.26$\,\mu$m 
at spectral and spatial resolutions of $\approx3.3\,\AA$ and 0\farcs15$\pm$0\farcs03, respectively. The total exposure time at each band was 
4800~s.

The final J and K$_{\rm l}$ data cubes, obtained by combining the individual exposures (which included some dithering), contain $\approx$4200 spectra 
at an angular sampling of $0\farcs05\,\times\,0\farcs05$, covering the central 700$\times$700 pc$^2$ of Mrk\,1066.
 For a detailed description of the observations and data reduction procedures see \p1.


\section{Results}

As shown in \p1,  
in the J band we observe the following emission lines: [P {\sc ii}]\,$\lambda$1.14713 
and 1.18861\,$\mu$m, [Fe\,{\sc ii}]\,$\lambda$1.25702, 1.27912, 1.29462, 1.29812, 1.32092 and 1.32814\,$\mu$m,  H\,{\sc i}~\pb, the He\,{\sc ii} line at 1.16296\,$\mu$m
 and the [S\,{\sc ix}] coronal line at 1.25235\,$\mu$m. 
 
In the \kl\ band we observe the H$_2$ lines at 2.12183, 2.15420, 2.22344, 2.24776, 2.40847, 
2.41367, 2.42180, 2.43697 and 2.45485\,$\mu$m, the  H\,{\sc i}~\br, the  He\,{\sc i}\,$\lambda$2.14999\,$\mu$m and the [Ca\,{\sc viii}]\,$\lambda$2.32204\,$\mu$m 
coronal line.  Besides the emission lines, the CO stellar absorption band heads around 2.3\,$\mu$m  are also present in the spectra an have been used to derive the 
 stellar kinematics.

\subsection{Stellar kinematics}\label{stel_kin}

\begin{figure*}
\centering
\includegraphics[scale=1.0]{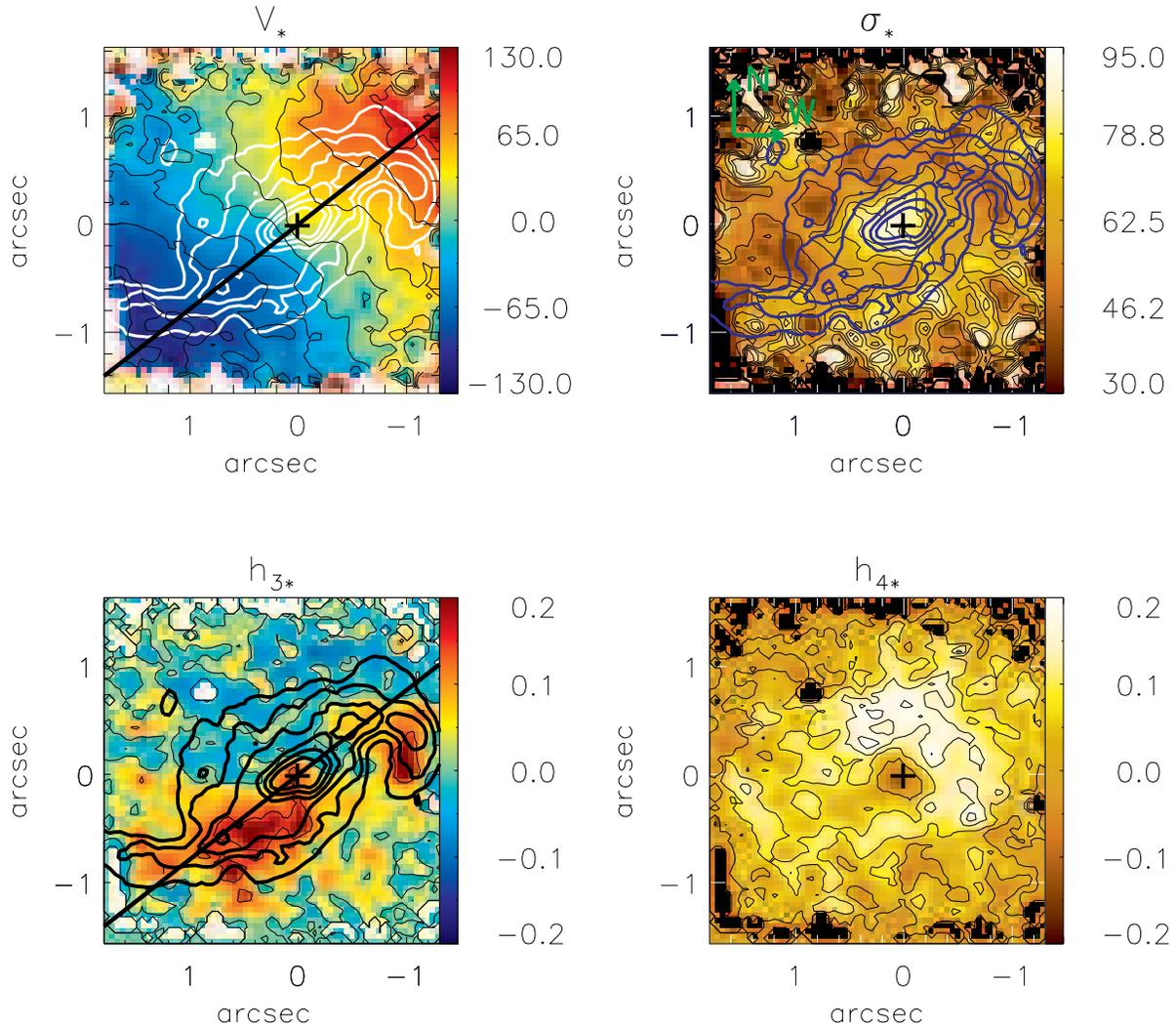}
\caption{Stellar kinematic maps obtained from the pPXF fit. Top: centroid velocity (left) and velocity dispersion (right) maps. Bottom: $h_{3*}$ (left) and $h_{4*}$ (right)
  Gauss-Hermite moments. The mean uncertainties are 6 km\,s$^{-1}$ for centroid velocity,  8 km\,s$^{-1}$ for $\sigma_*$, and 0.03 for $h_{3*}$ and $h_{4*}$. 
The continuous black line shows the orientation of the line of nodes obtained from the modeling of the velocity field. The thick contours overlaid to the velocity field, velocity dispersion and $h_{3*}$ maps are from the HST optical continuum image obtained with the WFPC2 through the filter F606W.}  
\label{stel}
\end{figure*}

In order to obtain the stellar line-of-sight velocity distributions (LOSVD) we fitted the $^{12}$CO and $^{13}$CO  stellar absorption band heads around 2.3\,$\mu$m in the $K$-band spectra using the penalized Pixel-Fitting (pPXF) method of \citet{cappellari04}. This method finds the best fit to a galaxy spectrum 
by convolving stellar template spectra with a given LOSVD [$L(v)$],
represented by Gauss-Hermite series \citep[e.g.][]{vandermarel93,gerhard93}:
\begin{equation}
L(v)=\frac{{\rm e}^{-(1/2)/y^2}}{\sigma_*\sqrt{2\pi}}\left[1+\sum_{m=3}^M h_{m*} H_m(y) \right],
\end{equation}
where $y = (v -V_*)/\sigma$, $V_*$ is the stellar centroid velocity, $\sigma_*$ is the velocity dispersion, $v=c{\rm\, ln}\lambda$ and $c$ is the speed of light. The $H_m$ are the Hermite polynomials, $h_{m*}$ are the Gauss-Hermite moments \citep{cappellari04}. The pPXF routine outputs the $V_*$, $\sigma_*$ and the higher-order Gauss-Hermite moments $h_{3*}$ and $h_{4*}$. As discussed in \citet{riffel08} and  \citet{winge09}, the use of a library of stellar templates, instead of a single stellar spectrum, is fundamental for a reliable derivation of the stellar kinematics. In this work we have used template spectra from the  Gemini library of late spectral type stars observed with the Gemini Near-Infrared Spectrograph (GNIRS) IFU and  NIFS \citep{winge09}.

In Figure\,\ref{stel_fit} we show some fits (in red) compared to the observed spectra (in black). This figure illustrate the fits at four positions: the nucleus (top-left panel), 0\farcs5~north-west (top-right panel), 
0\farcs5~south-east (bottom-left panel) and 0\farcs5~north-east (bottom-right panel).  
High signal-to-noise (S/N) ratio ($\gtrsim20$) is required to obtain a reliable fit of the observed spectra. 
As the S/N decreases with distance from the nucleus, we have replaced the flux at each spatial pixel
by the average flux of the 9 nearest pixels, which increased the S/N ratio and allowed us 
to measure the stellar kinematics at most locations, except for a few pixels close to the borders of the IFU field.
 This procedure can be justified by the fact that our data cube is somewhat oversampled (the pixels correspond to $0\farcs05\,\times\,0\farcs05$), and this average is thus equivalent to use an effective aperture of the 
order of the PSF FWHM ($\approx\,0\farcs15$).

In Figure\,\ref{stel} we present two-dimensional maps of the stellar
kinematics. The central cross marks the position of the nucleus, defined as the locus of the peak of the \kl~band continuum. 
The top left panel of Fig.\,\ref{stel} shows the stellar 
velocity field, from which we subtracted the systemic velocity, obtained 
from the modeling of this  velocity field (see Sec.~\ref{disc_stel}). The corresponding heliocentric value is  $V_s=3587\pm9\,{\rm km\,s^{-1}}$. The continuous black line shows the orientation 
of the  line of nodes  -- $\Psi_0=128^\circ\pm6^\circ$ measured from north to east, and obtained also from the modeling mentioned above. The mean
uncertainty in the centroid velocity is $\approx6\,{\rm km\,s^{-1}}$.  The velocity field shows blueshifts to the
 south-east and redshifts to the north-west and seems to be dominated by rotation, with an amplitude of $\approx$120\,\kms and 
 kinematical centre coincident with the position of the continuum peak, within the uncertainties.

In the top-right panel of Fig.~\ref{stel} we present the stellar velocity dispersion map, which has a mean uncertainty of 8~\kms~ and presents values 
ranging from 30 to 95~\kms, with the highest  ones observed at the nucleus.   A partial oval ring of low $\sigma_*$ values ($\approx50\,{\rm km\,s^{-1}}$) is observed surrounding the 
nucleus at $\approx 1^{\prime\prime}$ from it.  The bottom panels  show the higher order Gauss-Hermite moments $h_{3*}$ (left) and $h_{4*}$ (right).   
These moments measure deviations of the line profile from  a  Gaussian: the parameter $h_{3*}$ measures asymmetric deviations and the $h_{4*}$ measures symmetric 
deviations \citep[e.g.][]{vandermarel93}. The values $h_{3*}$ and $h_{4*}$ vary from $-$0.2 to 0.2 with mean uncertainty of 0.03.  The highest values of $h_{3*}$ 
are observed from the nucleus to $\approx$\,1\farcs3 south-east and at $\approx1^{\prime\prime}$ north-west from the nucleus, while the 
lowest values are observed over most of the north and north-east side of the galaxy.  The $h_{4*}$ map presents the lowest values at the nucleus, while the highest values 
are observed in a ring  with radius $\approx 1^{\prime\prime}$ approximately coincident with the ring of low $\sigma_*$ values.

\subsection{Gaseous kinematics}

The most commonly used method to measure the gaseous kinematics is the fit of the emission-line profiles by Gaussian curves. 
Nevertheless, the emission lines in Mrk\,1066 present asymmetric profiles at many locations, which cannot be well represented by Gaussian curves.
 We have thus used Gauss-Hermite series \citep[e.g.][]{vandermarel93,profit}, using the same formalism used to derive the stellar kinematics (see Section~\ref{stel_kin}) in order to obtain a better fit to the emission-line profiles and a more reliable measurement 
of the centroid velocity (from the centroid wavelength of the profile), velocity dispersion ($\sigma$), as well as to obtain
the higher order $h_3$ and $h_4$ moments. Each emission line profile was fitted by the following equation \citep{profit}:

 \begin{equation}
  GH=\frac{A\alpha(w)}{\sigma}\left[1 + h_3 H_3(w) + h_4 H_4(w)\right],
 \label{ghfit}
 \end{equation}
 where $w\equiv\frac{\lambda-\lambda_c}{\sigma}$  and $\alpha(w)=\frac{1}{\sqrt{2\pi}}e^{-w^2/2}$, $A$ is the amplitude of the Gauss-Hermite 
 series, $\lambda_c$ is the centroid wavelength, $h_3$ and $h_4$ are the Gauss-Hermite moments and $H_3(w)$ and $H_4(w)$ are Hermite polynomials.

%

We added a linear component to Eq.\,\ref{ghfit} in order to fit the continuum under each emission-line and thus
the resulting equation contains seven free parameters ($A,\, \lambda_c, \sigma,\,h_3\,h_4$ plus two parameters for the linear equation), which have been determined by fitting the line profiles. 
The fitting of the emission-line profile was done using 
 the {\sc mpfit}\footnote{The MPFIT routine can be obtained from the Markwardt IDL Library at http://cow.physics.wisc.edu/~craigm/idl/idl.html} routine, 
 written in IDL\footnote{http://www.ittvis.com} programing language using the Levenberg-Marquardt least-squares method, 
in which initial guesses are given for the free parameters. More details about the fitting procedure and its implementation can be found in \citet{profit}.

\begin{figure}
 \centering
 \begin{minipage}{1\linewidth}
 \includegraphics[scale=0.22]{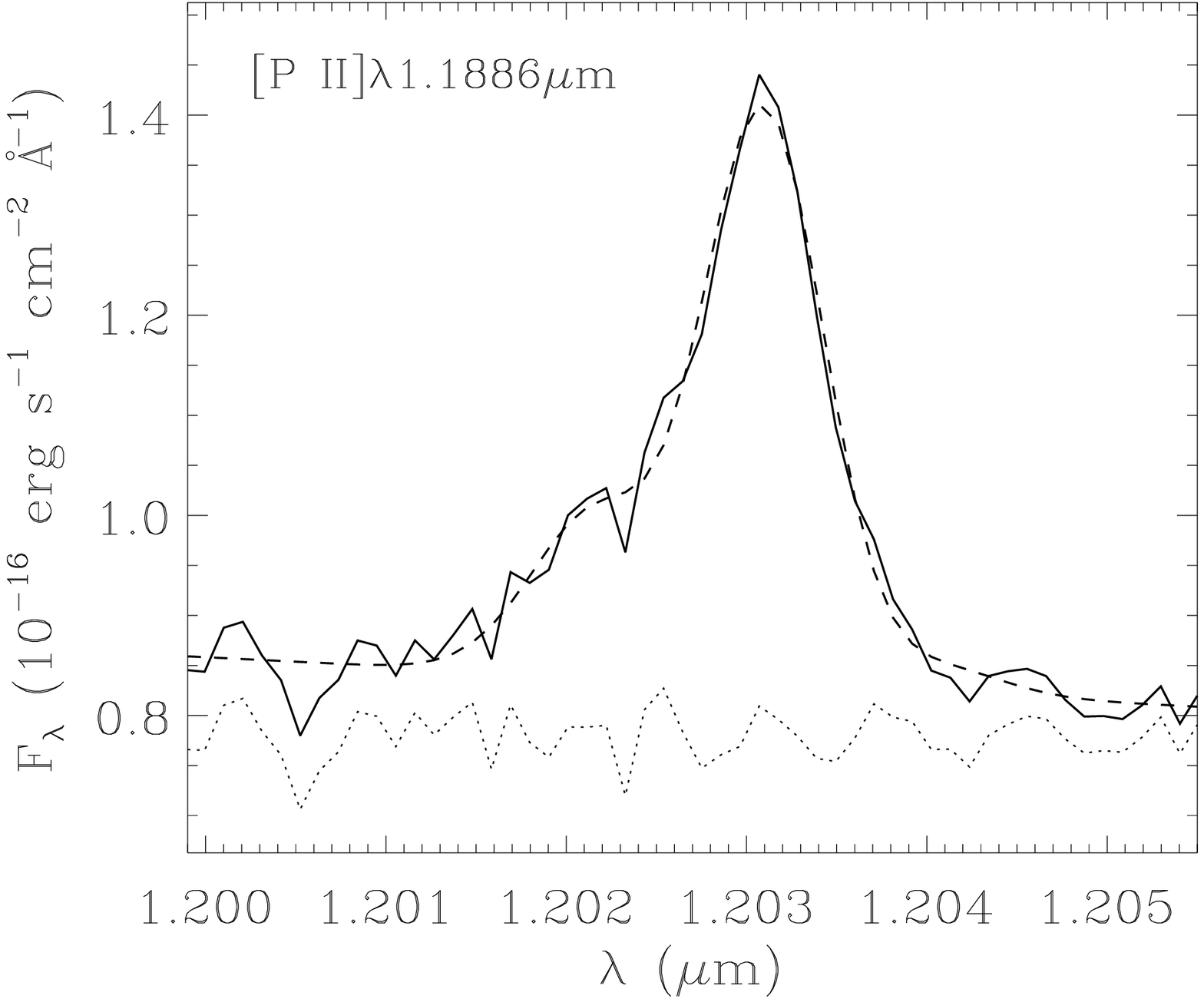}
 \includegraphics[scale=0.22]{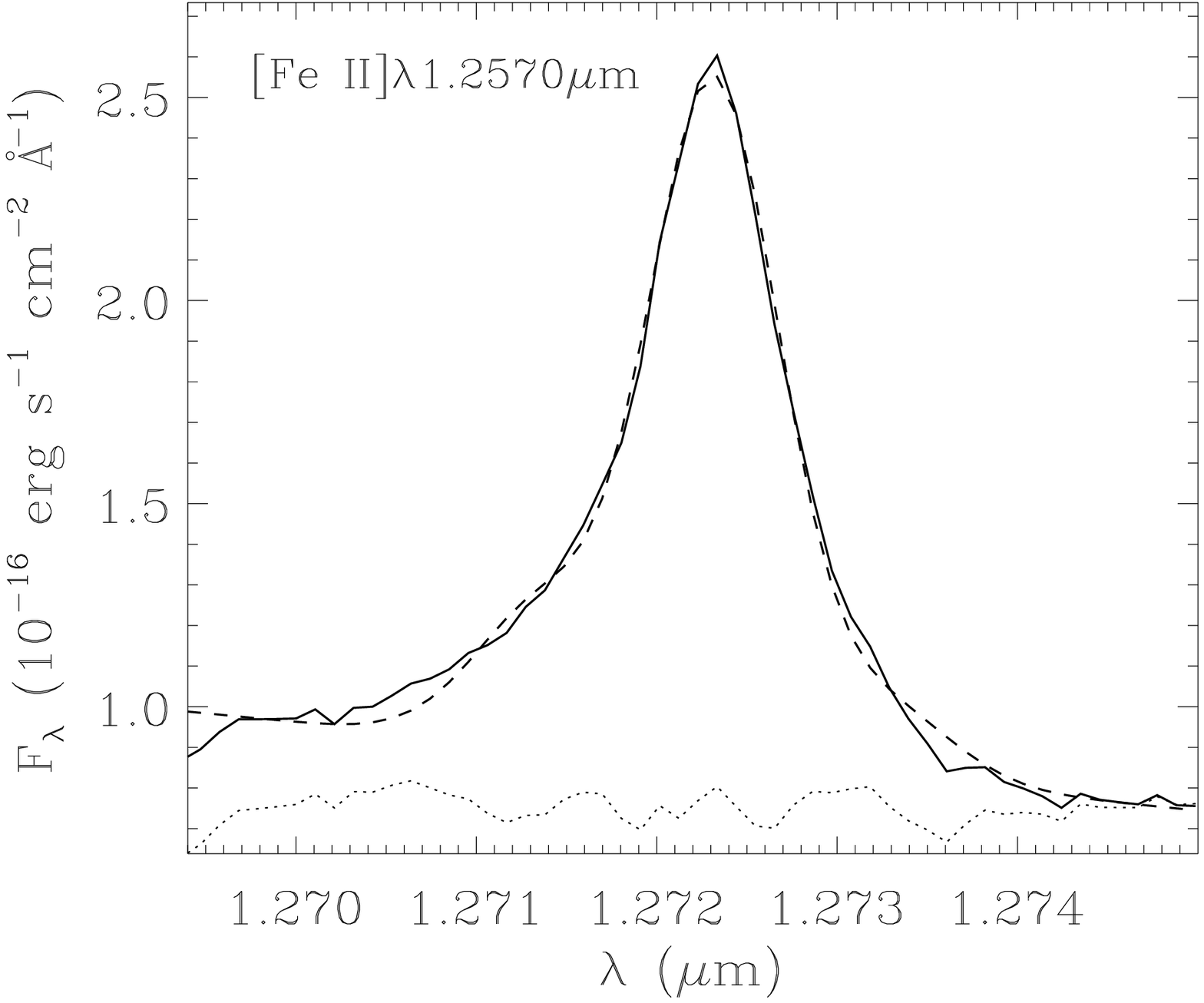}
 \end{minipage}
 \begin{minipage}{1\linewidth}
 \includegraphics[scale=0.22]{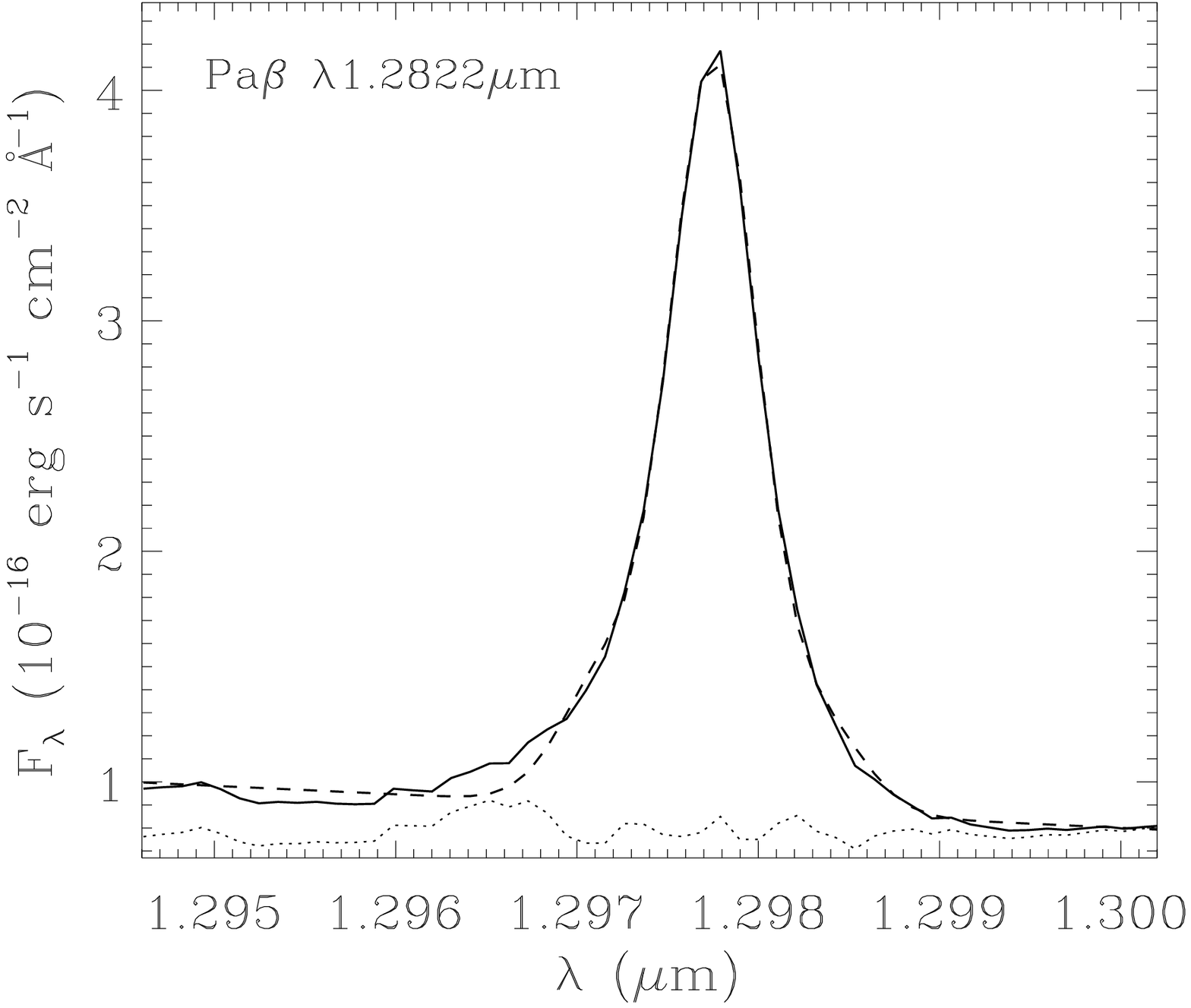}
 \includegraphics[scale=0.22]{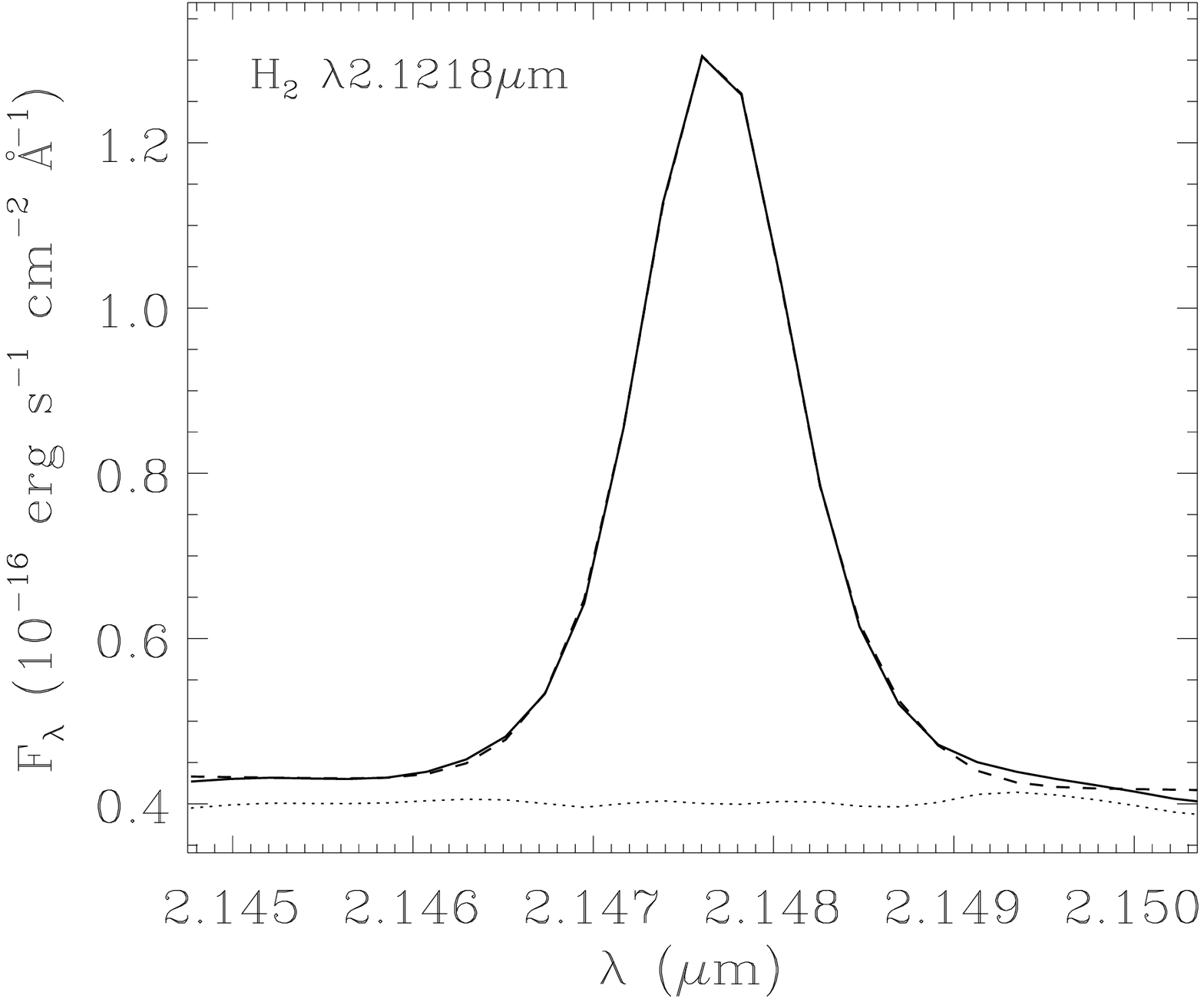}
 \end{minipage}
 \caption{Sample fits (dashed lines) of the observed emission-line profiles (solid lines) by Gauss-Hermite series  
 at 0\farcs5 north-west from the nucleus for an aperture of 0\farcs25$\times$0\farcs25. The dotted lines show the residuals of the fit plus an arbitrary constant (for visualization purposes only).
Top:[P\,{\sc ii}]\,$\lambda$1.1886\,$\mu$m (left) and [Fe\,{\sc ii}]\,$\lambda$1.2570\,$\mu$m (right) emission lines. Bottom: Pa$\beta$ and H$_2\,\lambda$2.1218$\mu$m emission lines.} 
 \label{fit}  
 \end{figure}

We have chosen the [P\,{\sc ii}]\,$\lambda$1.1886\,$\mu$m, 
 [Fe\,{\sc ii}]\,$\lambda$1.2570\,$\mu$m, Pa$\beta$ and H$_2\,\lambda$2.1218$\mu$m  emission lines to map the kinematics of the central
 region of Mrk\,1066. These 
lines trace the kinematics of distinct gaseous species (ionized gas forbidden lines, ionized gas permitted lines and molecular gas).
 These particular lines have been chosen in order to minimize the measurement uncertainties  because they have the highest S/N ratios among their species.
In Fig.\ref{fit} we show a sample of fits obtained for each line profile at 0\farcs5~north-west from the nucleus
 within an aperture of 0\farcs25$\times$0\farcs25. The solid line shows the observed profile, the dashed line shows
 the fit and the doted line shows the residuals (plus an arbitrary constant for visualization purposes). As shown in Fig.\,\ref{fit} the line profiles are well fitted by the Gauss-Hermite series and thus allow reliable determinations of  the centroid velocity, velocity dispersion, $h_3$  and $h_4$.

\begin{figure*}
\centering
\includegraphics[scale=1.0]{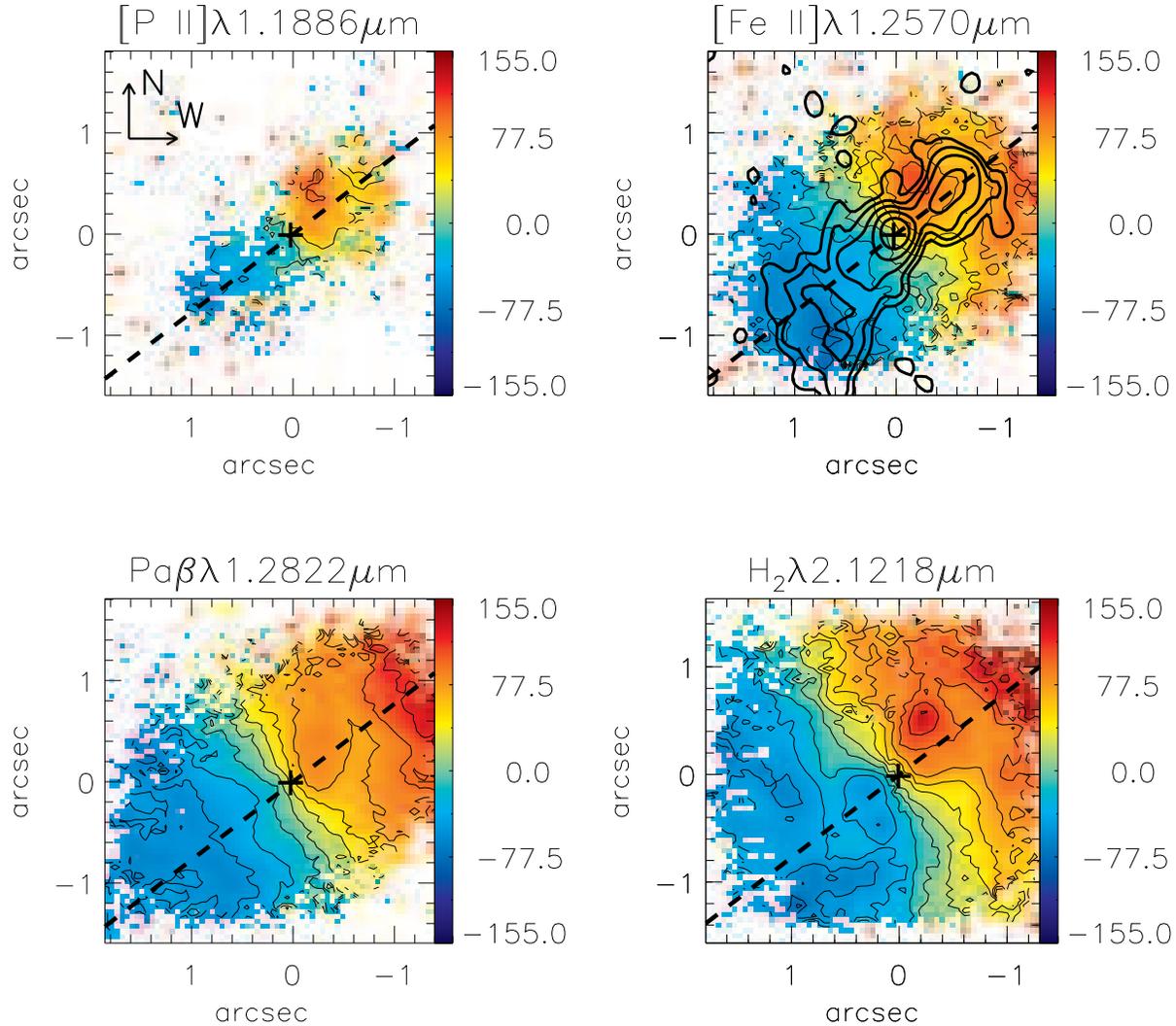}
\caption{Centroid velocity fields for the [P\,{\sc ii}]\,$\lambda$1.1886\,$\mu$m (top-left panel),  [Fe\,{\sc ii}]\,$\lambda$1.2570\,$\mu$m (top right), Pa$\beta$ (bottom left) and 
 H$_2\,\lambda$2.1218$\mu$m  (bottom right) emitting gas. The mean uncertainties are smaller than 7 km\,s$^{-1}$ for all emission lines. The thick black 
contours overlaid to the \feii\ velocity field are from a radio continuum image from \citet{nagar99}. 
The cross marks the position of the \kl~band continuum peak and the dashed line shows the orientation of the line  of nodes obtained from the modeling of the 
stellar velocity field. }  
\label{vel}
\end{figure*}

Figure~\ref{vel}  shows the velocity fields obtained from the Gauss-Hermite fit of the  profiles for the  [P\,{\sc ii}]\,$\lambda$1.1886\,$\mu$m, [Fe\,{\sc ii}]\,$\lambda$1.2570\,$\mu$m, Pa$\beta$ and 
 H$_2\,\lambda$2.1218$\mu$m emission lines, after subtraction of the systemic velocity of the galaxy ($V_s=3587\,{\rm km\,s^{-1}}$). The crosses mark the 
position of the nucleus and the dashed lines show the orientation of the line of nodes derived from the modeling of the stellar velocity field (see Sec.~\ref{disc_stel}). The gas velocity fields are similar to each other and are dominated by rotation, with the south-east side approaching 
and the north-west side receding from us. Although similar to the stellar velocity field, the gas velocity fields are more disturbed.
 The most conspicuous feature is more clearly observed in the \h2\ velocity field: a gradient from $\approx$125~\kms~ at $\approx$0\farcs5 
north-west from the nucleus to $\approx-$55~\kms~ at $\approx$0\farcs5 south-east from it along the PA$\approx$150$^\circ$,  which 
seems to be a compact rotating disc immersed in the larger disc (stellar and gaseous).

\begin{figure*}
\centering
\includegraphics[scale=1.0]{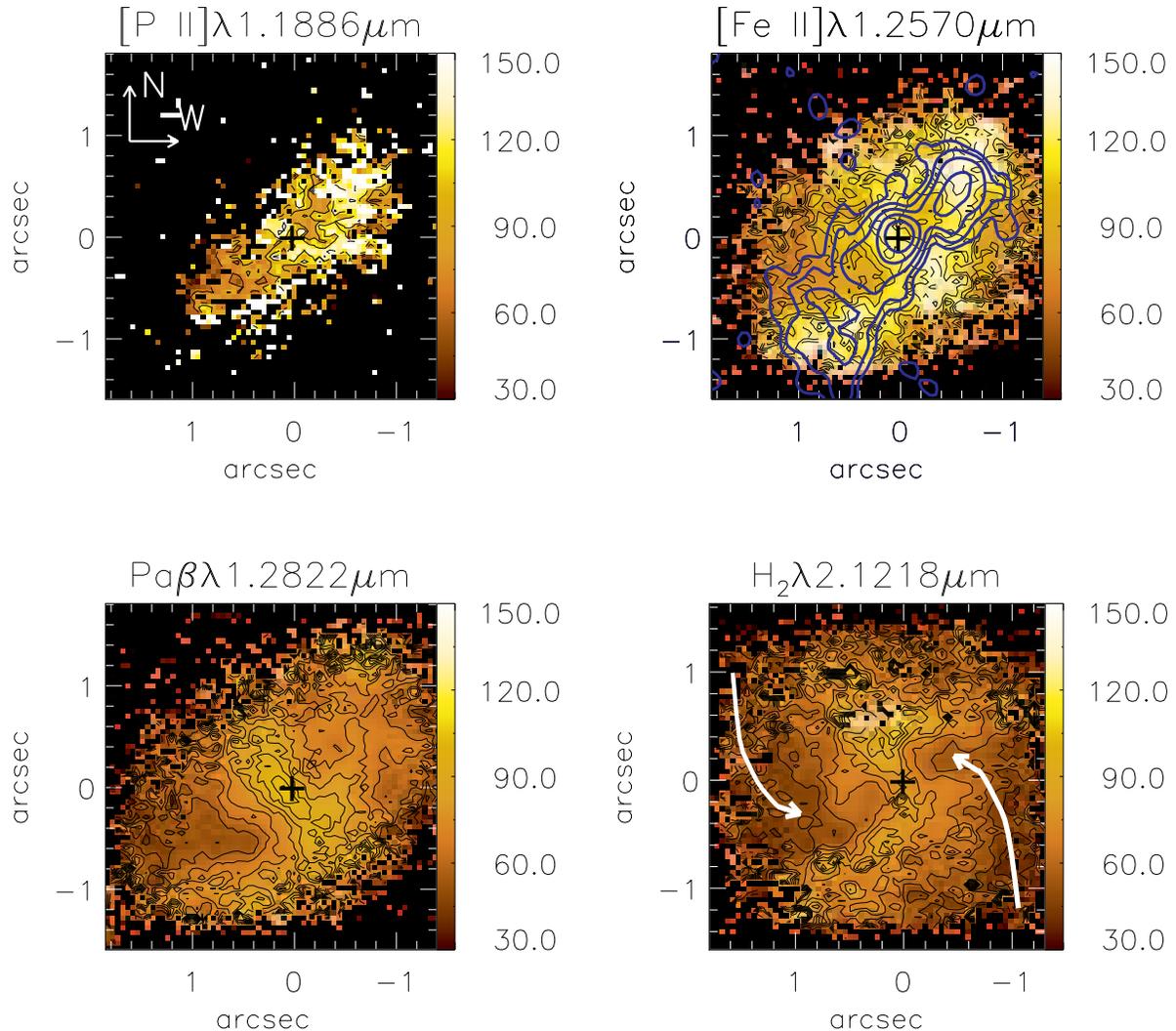}
\caption{Velocity dispersion maps for the [P\,{\sc ii}]\,$\lambda$1.1886\,$\mu$m (top-left panel),  [Fe\,{\sc ii}]\,$\lambda$1.2570\,$\mu$m (top right), Pa$\beta$ (bottom left) and 
 H$_2\,\lambda$2.1218$\mu$m  (bottom right) emission lines. The mean uncertainties are smaller than 10 km\,s$^{-1}$ for all emission lines.
The cross marks the position of the \kl~band continuum peak, the arrows identify the regions of asymmetry in the flux distribution, 
and blue contours overlaid on the \feii~$\sigma$ map are from the radio continuum map of \citet{nagar99}.}  
\label{sig}
\end{figure*}

In Figure~\ref{sig} we present the gas velocity dispersion ($\sigma$) maps. In order to compare our $\sigma$ maps with  the radio structure we have overlaid the contours (blue lines) of the 3.6~cm radio continuum image from \citet{nagar99}  on  the \feii~$\sigma$ map. The \feii~and \pii~$\sigma$ maps are similar and show the highest values of up to $\approx$150~\kms~in regions at and around the radio structure, while the lowest values ($\approx50$~\kms) are observed  predominantly to south-east of the nucleus.  The \pb~$\sigma$ map presents smaller values than those of  \feii~and \pii, with the highest values of $\approx$100~\kms\ observed approximately along the minor axis of the galaxy. The lowest values ($\le$\,60\,\kms) are observed in two structures resembling spiral arms, more clearly observed in the  \h2~$\sigma$ map. This map also shows that these arms -- one coming from  the east and north-east and the other coming from  the west and south-west -- seem to curve into the major axis towards the centre into a structure which we have identified with a compact nuclear disc in Sec.\,\ref{gas_kin}.

\begin{figure*}
\centering
\includegraphics[scale=0.7]{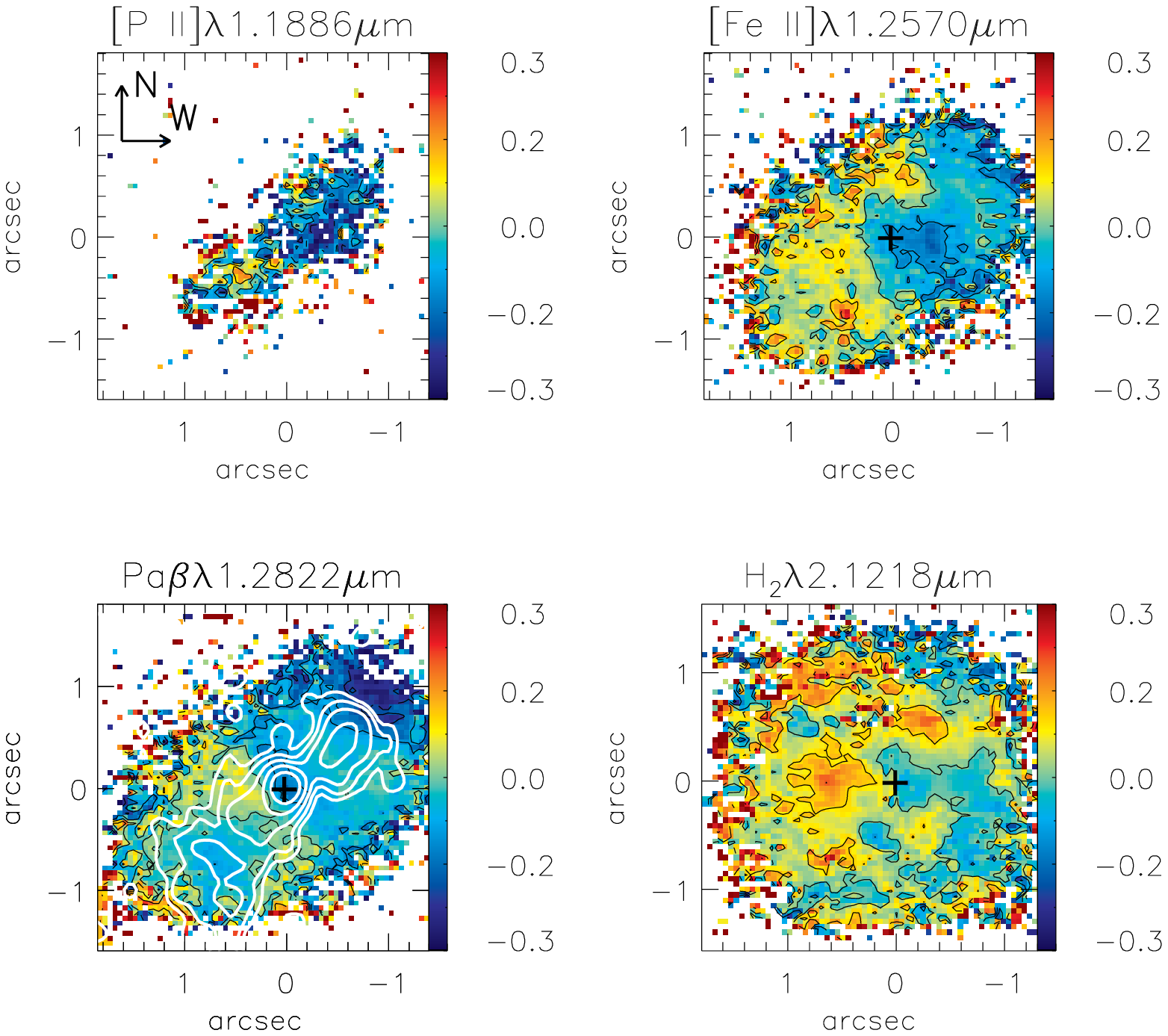}
\caption{$h_3$ Gauss-Hermite moment maps for [P\,{\sc ii}],  [Fe\,{\sc ii}], Pa$\beta$  and 
 H$_2$ emission lines. The white contours overlaid to the Pa\,$\beta$~$\sigma$ map are from the radio continuum image of \citet{nagar99}.}  
\label{h3}
\end{figure*}

\begin{figure*}
\centering
\includegraphics[scale=0.7]{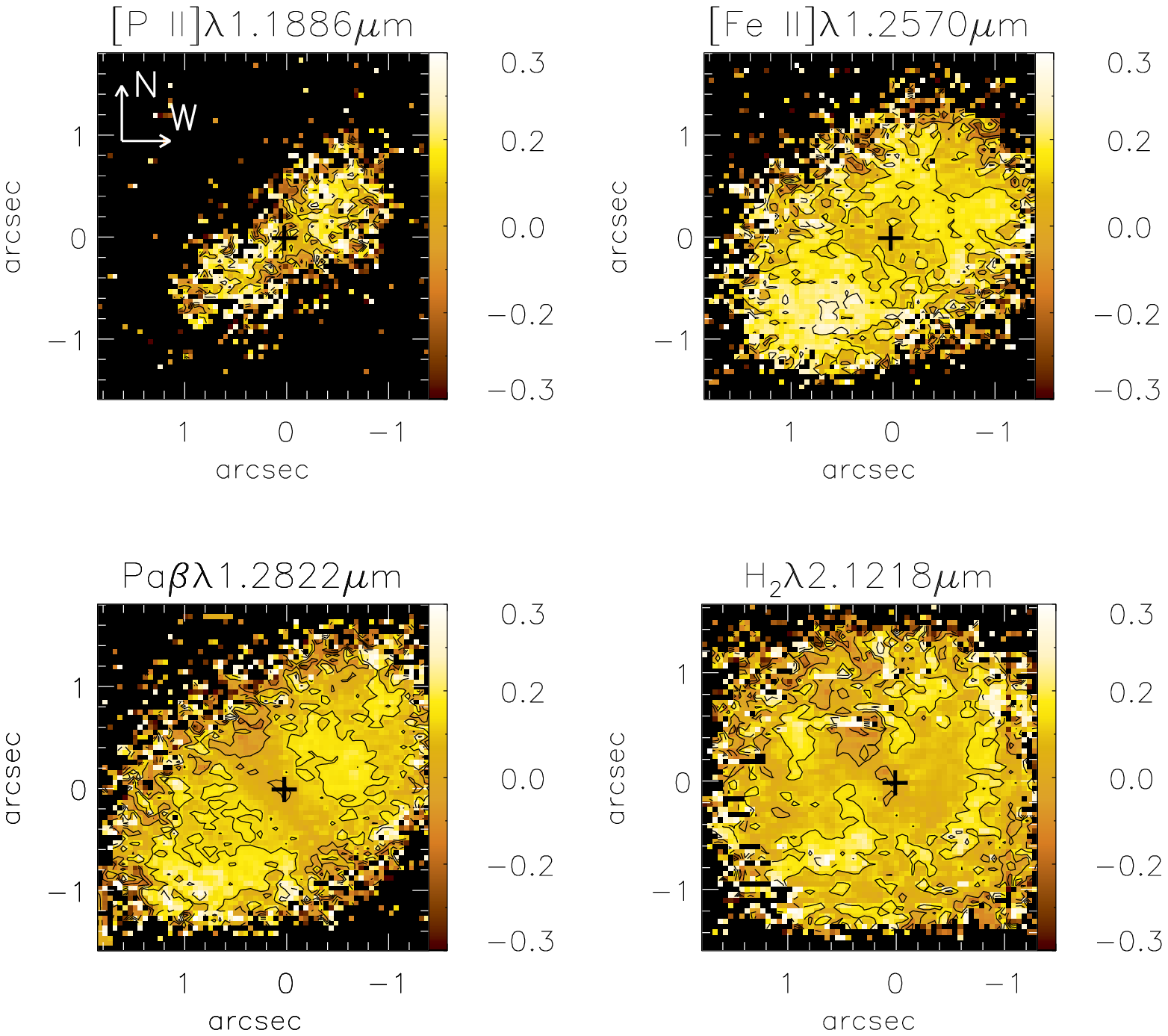}
\caption{Same as Fig.~\ref{h3} for the  $h_4$ Gauss-Hermite moment.}
\label{h4}
\end{figure*}

In Figures~\ref{h3} and \ref{h4} we present the $h_3$ and $h_4$ Gauss-Hermite moment maps. 
For \pii and \feii, the $h_3$ maps show negative values to north-west of the nucleus, while the \pb\ presents negative $h_3$ 
values at most locations (except to east and south of the nucleus). However the most negative values reaching $\approx -0.3$ are observed predominately at
 $\sim$\,1\arcsec  north-west of the nucleus at the edge of the radio jet, indicating the presence of a blue wing in the emission-line profiles
 at this location.  Some regions with positive $h_3$, with values of $\approx$\,0.1 and reaching 0.2 at a few locations, are observed to the north, north-east, east and south-east of the nucleus for \feii and H$_2$, indicating the presence of a red wing in the emission-line profiles at these locations.
 
The $h_4$ maps present values close to zero at and around the nucleus for all emission lines -- indicating no significant deviations from a Gaussian in the emission-line profiles. 
At $\sim$\,0\farcs5 from the nucleus 
approximately along the major axis, the $h_4$ maps for the ionized gas show regions with values $h_4\approx$0.15, which indicate that the profiles 
are narrower and have broader wings than Gaussians at these locations.


Coronal lines are also present in the spectra of Mrk\,1066, such as [S\,{\sc ix}]\,$\lambda1.25235\,\mu$m and [Ca\,{\sc viii}]\,$\lambda$2.32204\,$\mu$m. The  \caviii\ and [S\,{\sc ix}]\ ions have 
ionization potentials of 127.2~eV and 328.8~eV, respectively \citep{storchi-bergmann09} and their emission lines 
can be used to map the kinematics of the high ionization emitting gas. Nevertheless, as discussed in \p1, their flux distributions 
are unresolved by our observations. The \caviii\ line is affected by the $^{12}$CO\,(3-1)\,$\lambda$2.323$\mu$m absorption and we have thus subtracted the stellar population (obtained
 from the fit of the stellar kinematics) from the galaxy spectra before fitting the  [Ca\,{\sc viii}] emission-line profile. The fit gave
centroid velocity  $V\approx-$12~\kms, which is consistent with the systemic velocity of the galaxy within the uncertainties, and velocity dispersion of  $\sigma\approx$145~\kms. The fit of the [S\,{\sc ix}]\,$\lambda1.25702\,\mu$m  emission line gives  $V\approx$290~\kms\ and $\sigma\approx$70~\kms, but the uncertainties are large because the  [S\,{\sc ix}] line profile is contaminated by both the [Fe\,{\sc ii}]\,$\lambda$1.2570\,$\mu$m emission  line and  underlying stellar absorptions.

\subsection{Channel maps}\label{chamaps}

\begin{figure*}
\centering
\includegraphics[scale=0.8]{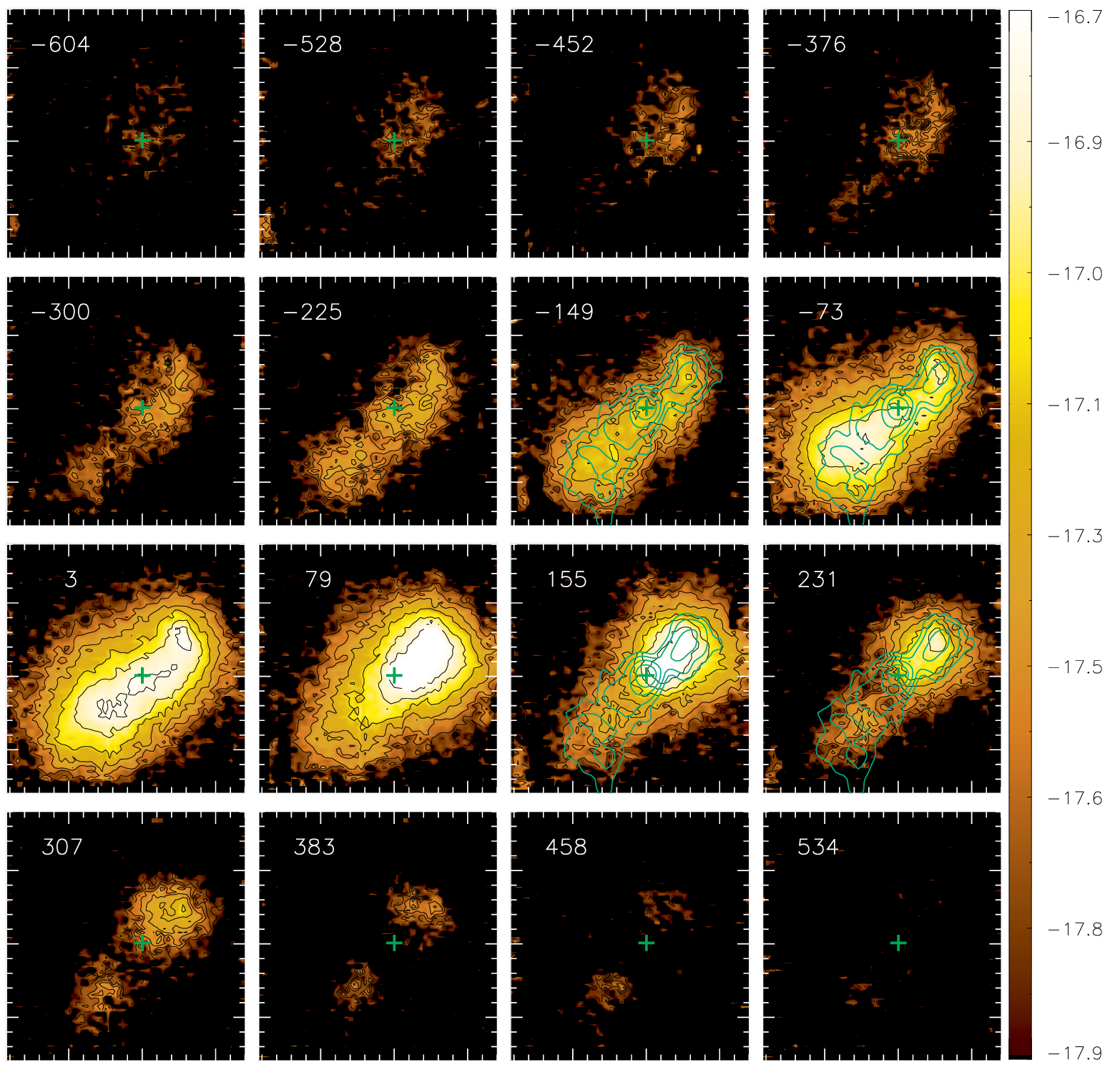}
\caption{Channel maps along the \feii$\,\lambda1.2570\,\mu$m emission-line profile. Each panel shows the intensity distribution at the velocity shown in the 
top of the panel, integrated within a velocity bin of $\approx$75~\kms. The cross marks the position of the nucleus and the green contours overlaid to some 
panels are from the 3.6~cm radio image.}
\label{slice_feii}
\end{figure*}

In order to better sample the gas kinematics over the whole velocity distribution, including the highest velocities in the wings of the line profiles,  we ``sliced'' each profile into a sequence of velocity channels. 
In Figure~\ref{slice_feii} we present the channel maps for the \feii$\,\lambda1.2570\,\mu$m emission line for a velocity bin of $\approx$75~\kms, corresponding to three spectral pixels. Each panel presents the flux distribution in logarithmic units integrated within the velocity bin and  centred at the velocity shown in the top-left corner of the panel (relative to the systemic velocity of the galaxy). The central cross marks the position of the nucleus and the green contours
 overlaid to some panels are from the 3.6~cm radio image of \citet{nagar99}. The 
highest blueshifts of up to $-$500~\kms\ are observed to north-west of the nucleus at $\approx$0\farcs85 from it, approximately coincident with  a hot spot observed in the radio map. When the blueshifts reach smaller absolute values, of $\approx\,-$300~\kms, the flux distribution  presents emission to  both sides of the nucleus along the PA$\approx$135/315$^\circ$, which is approximately the PA of both the radio jet and major axis of the galaxy. 
The highest redshifts, of up to  of $\approx$450~\kms, have origin in gas located at both sides of the 
nucleus at distances of $\approx$0\farcs85 to north-west and  $\approx1^{\prime\prime}$ to south-east.

\begin{figure*}
\centering
\includegraphics[scale=0.8]{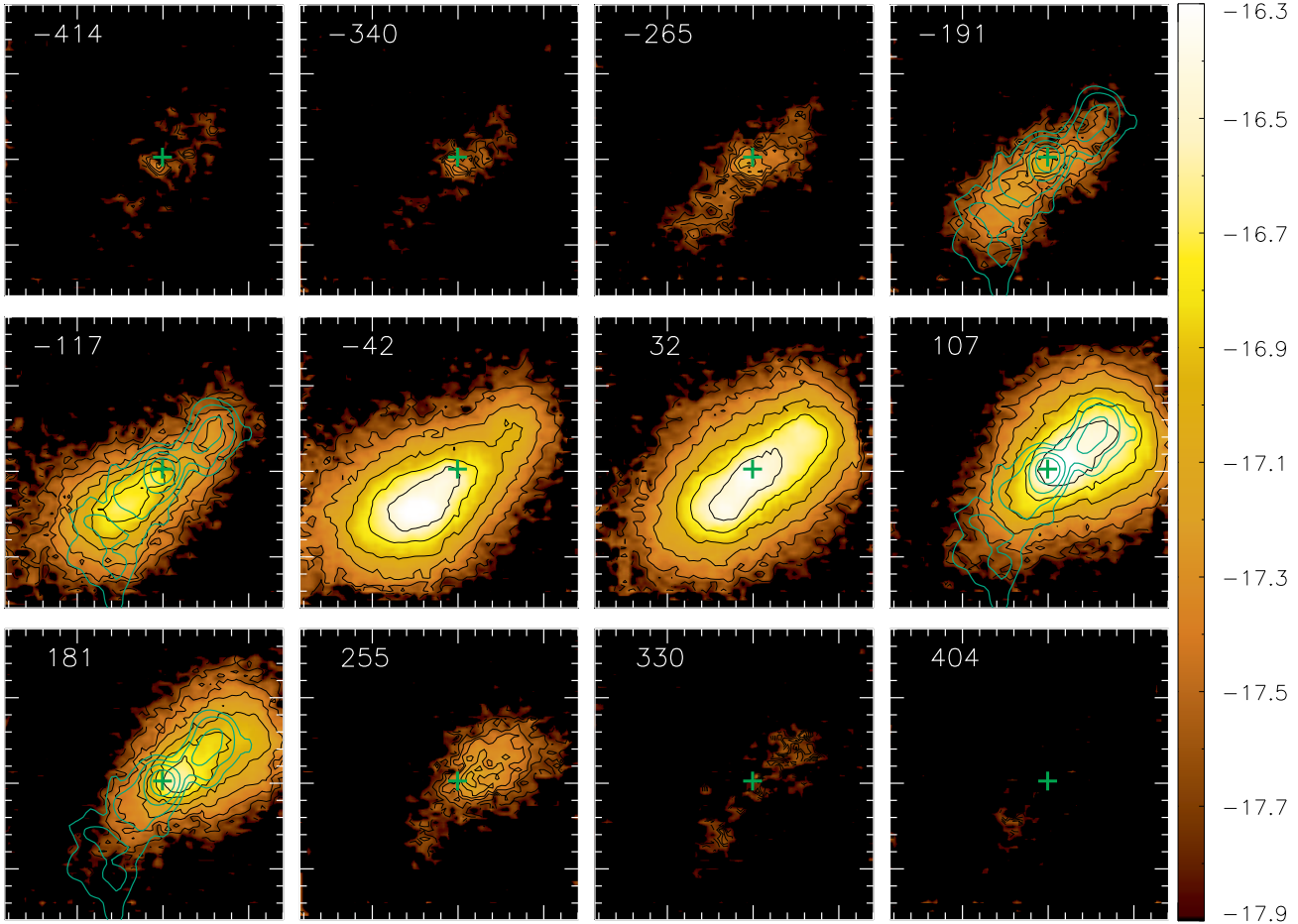}
\caption{Same as Fig.~\ref{slice_feii} for the \pb\ emission-line profile.}
\label{slice_pb}
\end{figure*}

Figure~\ref{slice_pb} presents the velocity channels for the \pb\ emission line. The highest \pb\ velocities -- 
blueshifts of $-$350~\kms\ and redshifts of 320~\kms -- are smaller than those observed for \feii. At the highest blueshifts 
 the emission is dominated 
by the nucleus, showing fainter emission from the north-west and south-east sides, 
while most of the redshifts are due to emission from the north-west side of the galaxy. 
From blueshifts of $\approx\,-$100\,\kms\ to redshifts of $\approx\,$100\,\kms\ the peak of the emission shifts 
from the south-east to the north-west of the nucleus along the galaxy major axis, suggesting rotation.

\begin{figure*}
\centering
\includegraphics[scale=0.8]{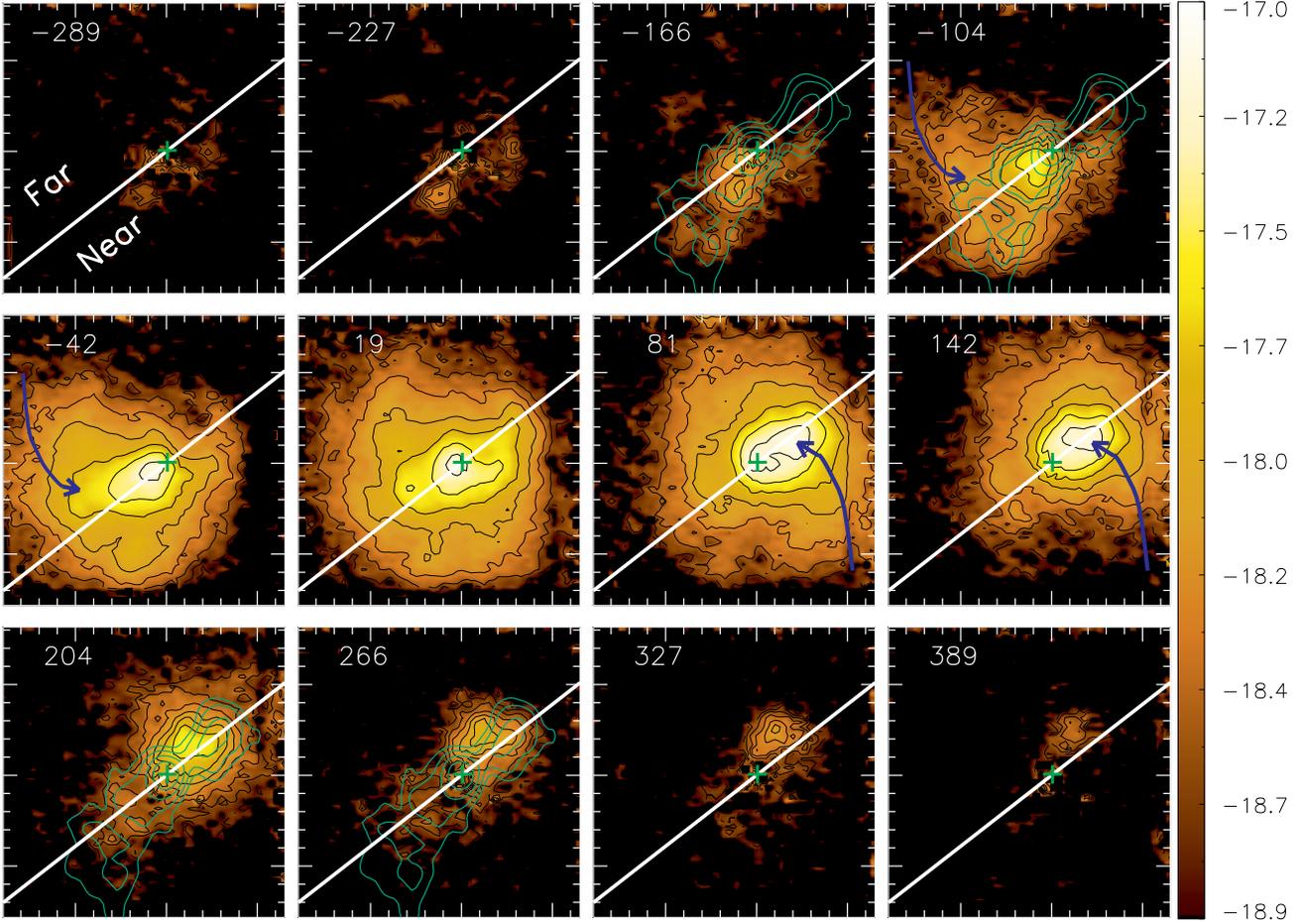}
\caption{Velocity slices along the \h2$\,\lambda2.1218\,\mu$m emission-line profile. Each panel shows the intensity distribution at the velocity shown in the top-left corner of the panel, integrated within a velocity bin of $\sim$60~\kms. The cross marks the position of the nucleus, the white line shows the galaxy major axis, the arrows identify regions {where we identify an asymmetry in the flux distribution relative to the major axis}, and the green contours overlaid to some panels are from the 3.6~cm radio image.}
\label{slice_h2}
\end{figure*}

The velocity channel maps for the \h2$\,\lambda2.1218\,\mu$m emission line are shown in Figure~\ref{slice_h2} for velocity bins of $\sim$60~\kms, corresponding to two spectral pixels. The pattern here is somewhat distinct from those observed for \feii\ and \pb.  The \h2\ emission is more uniformly distributed over the whole IFU field, being less ``collimated'' along the radio axis, with  blueshifts  observed mostly to the south-east and redshifts to the north-west. This is what is expected if the velocity field is dominated by rotation similar to that of the stars. Nevertheless, it can be noticed that the velocity fields are not symmetric relative to the galaxy  major axis (what should be the case for pure rotation). For example, in the blueshifted channels centred at $-104$\,\kms\ and $-42$\,\kms\ the flux distribution curves towards the north, while in the redshifted channels centred at 81\,\kms\ and 142\,\kms\ the flux distribution suggest some curvature to the south. These regions are marked with arrows in Figure~\ref{slice_h2}. 

The \pii$\,\lambda$1.1886\,$\mu$m velocity-channel maps are similar to those for \feii, but are much noisier and are thus not shown here.

\section{Discussion}

\subsection{Stellar kinematics} \label{disc_stel}

As can be observed in Fig.~\ref{stel}, the stellar velocity field presents a rotation pattern, and  in order to obtain the systemic velocity, orientation of the line of nodes and an estimate for the bulge mass of Mrk\,1066 we have fitted the stellar velocities with a model of circular orbits in a plane subject to a Plummer gravitational potential, given by

 \begin{equation}
\Phi=-\frac{GM}{\sqrt{r^2+a^2}},
\end{equation}
where  $a$ is a scale length, $r$ is the distance from the nucleus in the plane of the galaxy, $M$ is the mass inside $r$ and $G$ is the Newton's gravitational constant. Although
 this model is very simple, it has approximately reproduced the stellar velocities in the central regions of other Seyfert galaxies in previous studies \citep{barbosa06,riffel08}.
  
Following \citet{barbosa06}, defining the coordinates of the kinematical centre  as ($X_0,Y_0$) relative to the photometric nucleus,
 the observed centroid velocity at position ($R,\Psi$), where $R$ is the projected  distance from the nucleus in the plane of the sky and 
$\Psi$ is the corresponding position angle, can be expressed as

\begin{equation}
V_r=V_s + \sqrt{\frac{R^2GM}{(R^2+A^2)^{3/2}}}\frac{{\rm sin}(i){\rm cos}(\Psi-\Psi_0)}{\left({\rm cos^2}(\Psi - \Psi_0) + \frac{{\rm sin^2}(\Psi-\Psi_0)}{{\rm cos^2}(i)} \right)^{3/4}}
\end{equation}
where $V_s$ is the systemic velocity, $i$ is the inclination of the disc ($i=0$ for a face on disc) and $\Psi_0$ is the position angle of
the line of nodes. The relations between $r$ and $R$, and between $a$ and $A$ are: $r=\alpha R$ and $a=\alpha A$, where 
$\alpha=\sqrt{{\rm{cos^2}(\Psi-\Psi_0)+\frac{{\rm sin^2}(\Psi-\Psi_0)}{{\rm cos}(i)^2}}}$.

\begin{figure}
\centering
\includegraphics[scale=0.9]{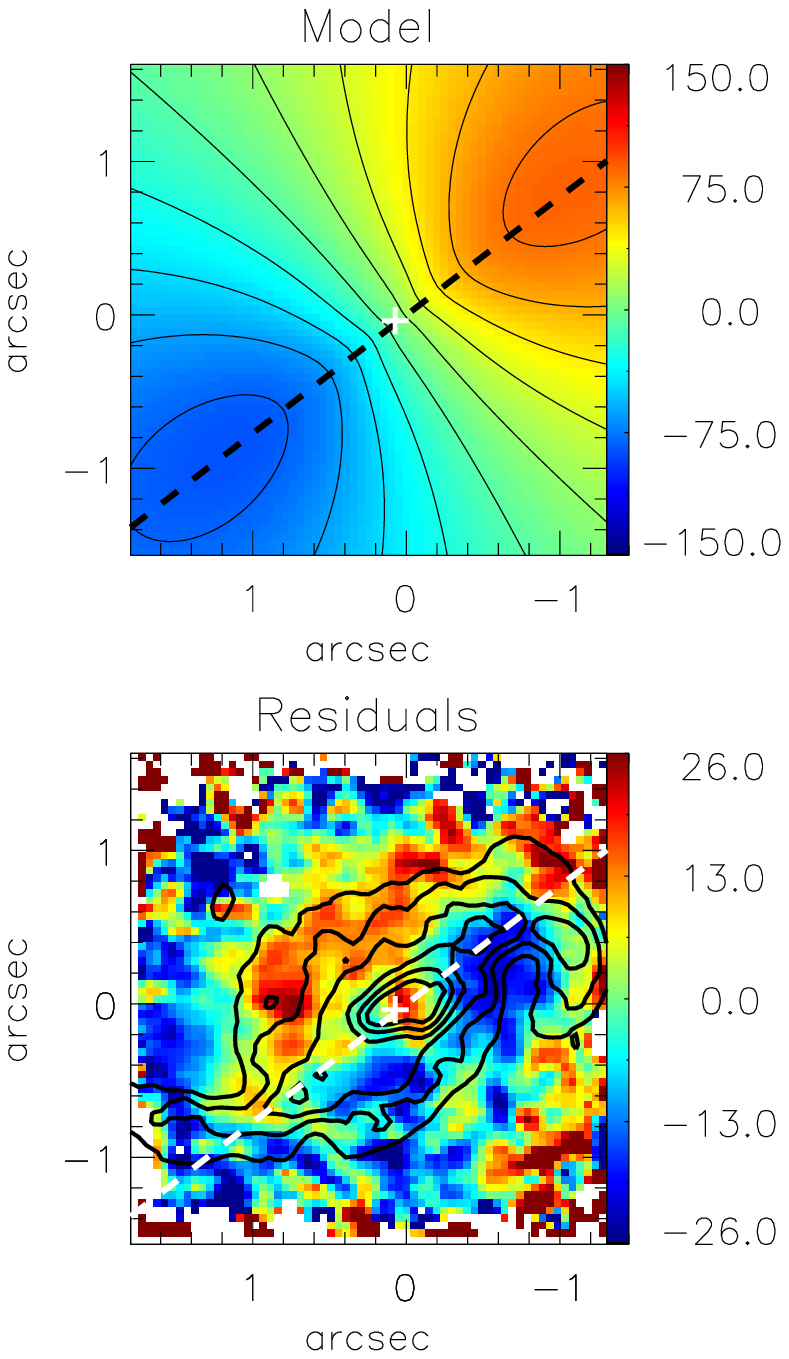}
\caption{Rotating disc model for the stellar velocity field of Mrk\,1066 (top panel) and residual map between the observed and modeled velocities (bottom panel). The  dashed line shows the orientation of the line of nodes ($\Psi_0=$128$^\circ$), the cross marks the position of the nucleus and the contours are from the HST optical image.}
\label{plummer}
\end{figure}

The equation above
contains six free parameters, including the kinematical centre, which
can be determined by fitting the model to the observed velocity field. This was
done using a Levenberg-Marquardt least-squares fitting algorithm, in which initial
guesses are given for the free parameters. In the fitting we considered a typical 
uncertainty in the observed velocity of 6\,\kms.

The parameters derived from the fit are: the systemic velocity corrected to the heliocentric reference frame $V_s=$3587\,$\pm$\,9\,km\,s$^{-1}$, 
$\Psi_0=$128$^\circ$\,$\pm$\,6$^\circ$, $M=$2.7$\pm$0.3\,$\times$\,10$^9$\,M$_\odot$, $i=38\pm5^\circ$,
and $A=$238\,$\pm$\,15\,pc.  The derived kinematical centre is coincident with the adopted position for the nucleus
(peak of the continuum emission) within the uncertainties, with $X_0=-$17$\pm$7\,pc and $Y_0$=3$\pm$3\,pc, considering that  our spatial resolution is (FWHM) $\approx$0\farcs15  (35\,pc). The uncertainty in the parameters were obtained directly from the application of the
Levenberg-Marquardt method, taking into account the uncertainty in  the velocity.


The  parameters derived from the stellar kinematics can be compared with those from previous works.  The systemic velocity derived here is about 40~\kms\ smaller than the one obtained by \citet{bower95} from optical emission lines and in agreement with the one derived by \citet{knop01} using near-IR emission lines. 
We consider our measurement of $V_s$  more robust than those from
previous works  as our observations have higher spatial resolution and $V_s$ was obtained from the stellar kinematics, which is a better tracer of the gravitational potential of the galaxy than the gas. 
The orientation of the line of nodes $\Psi_0=$128$^\circ$ is in reasonable agreement with the one obtained by \citet{knop01} -- $\Psi_0=$120$^\circ$.
The scale length $A$ and the bulge mass $M$  are similar to those obtained for the central region of other Seyfert galaxies using similar modeling \citep[e.g.][]{barbosa06}. 
It should be noticed that $M$ is tightly coupled with the inclination ($i$) of the disc, since $V^2 \propto M {\rm sin(}i{\rm )}$. Thus the uncertainties in $M$ and $i$ can be even higher than those quoted above due to this degeneracy. Nevertheless, the value obtained  for the inclination of the disc is in reasonable agreement with the one obtained by \citet{bower95}, $i=45^\circ$.

\begin{figure*}
\centering
\includegraphics[scale=0.8]{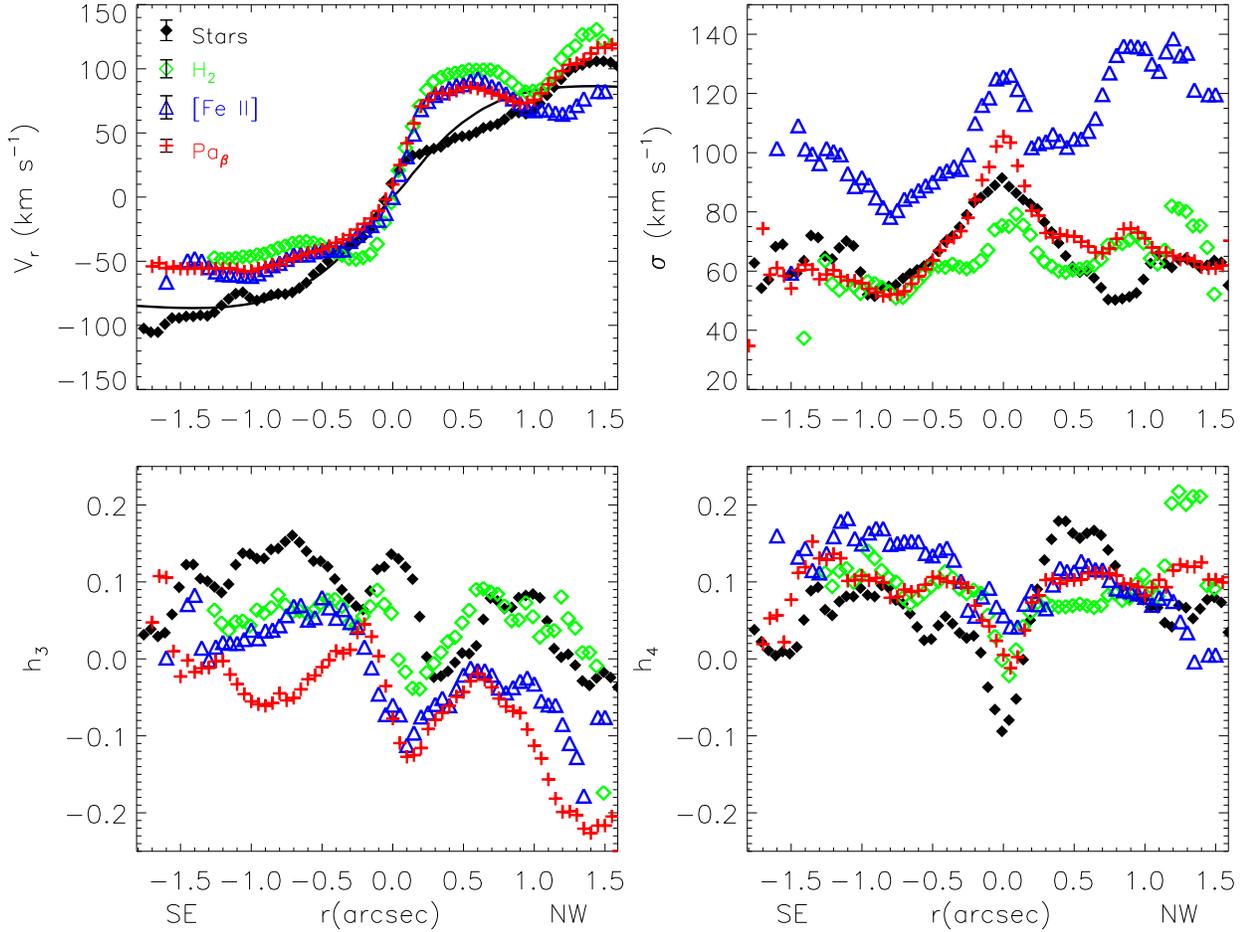}
\caption{One-dimensional cuts for the stellar and gaseous kinematics obtained along PA$=128^\circ$ (orientation of the line of nodes) within a pseudo-slit 0\farcs25 wide. Top: Radial velocities (left) and velocity dispersions (right). Bottom: $h_3$ (left) and $h_4$ (right) Gauss-Hermite moments. The model is shown as a continuous line in the top-left panel. Typical uncertainties for $\sigma$ are $\le$10~\kms\ and for $h_3$ and $h_4$ are $\approx$\,0.03.}
\label{cut_kin}
\end{figure*}

In figure~\ref{plummer} we show the best fit model for the stellar velocity field (top panel) and the residual map,  obtained from the subtraction of the model from the observed velocities. A comparison between the modeled and the observed velocities is also shown in the top-left panel of  Fig.~\ref{cut_kin}, where we  present one-dimensional cuts for the velocities along the line of nodes, averaged within a pseudo-slit  0\farcs25 wide.
The bottom panel of figure~\ref{plummer} shows that  the residuals between the observed  and modeled stellar velocity fields are usually smaller than 20~\kms ($\approx$\,15\% of the maximum velocity amplitude of the rotation curve), but present a systematic behaviour.  These systematic residuals -- blueshifts to the south-west of the major axis and redshifts to the north-east in the inner arc second  radius -- seem to correlate with the oval seen in the HST F606W optical image, as shown by the corresponding superimposed contours. A signature of the effect of this oval is already present in Fig.~\ref{stel}, where the ``zero-velocity curve''  shown shows an {\it S} shape, a known signature of the presence of a non-axisymetric structure in the gravitational potential, which is not present in  the model stellar velocity field of Fig.\,\ref{plummer} which shows straight isovelocity lines for velocities close to zero (along the kinematic minor axis).  
\citet{emsellem06} have found a similar S-shaped structure in the stellar velocity field of the Seyfert galaxy NGC\,1068, attributed  to the presence of a nuclear bar. In the case of Mrk\,1066, we attribute the distortions to the nuclear oval seen in the
the HST image  from which spiral arms seem to originate along PA=128$^\circ$ (see \p1). 


This oval is also seen in the stellar velocity dispersion map (top left panel of Fig.~\ref{stel}), in the form of a partial circumnuclear ring of low values with $\sigma_*\approx50\,{\rm km\,s^{-1}}$, at a distance of $\approx 1^{\prime\prime}$ from the nucleus, immersed in a  background with values of $\sigma_*\approx90{\rm\,km\,s^{-1}}$. This partial ring is also observed in the one-dimensional cut of the  $\sigma_*$ map shown in the  top-right panel of Fig.~\ref{cut_kin}. Similar circumnuclear rings of low velocity dispersions have been reported for several other Seyfert  galaxies \citep{barbosa06,riffel08,riffel09}, interpreted as colder regions with more recent star formation than the underlying bulge. 
This interpretation is supported by a recent study of the stellar population of this region \citep{paper2}, which shows a dominance of 
intermediate age ($10^8-10^9$\,yr) stars. 



The $h_{4*}$ map (bottom-right panel of Fig.~\ref{stel}) shows a ring of high values co-spatial with the ring of low $\sigma_*$ values.
Such $h_{4*}$ enhancement (due to a velocity distribution which is narrower than a Gaussian for low velocities) can be attributed to 
the contribution of the youngest stars (the stars actually span a range of intermediate ages) in the ring, which have a colder 
kinematics than the surrounding bulge stars.


The mass of the super-massive black hole in the centre of Mrk\,1066 can be estimated from the bulge stellar velocity dispersion using ${\rm log}(M_{BH}/{\rm M_\odot})=\alpha+\beta\,{\rm log}(\sigma_*/\sigma_0)$, where $\alpha=8.13\pm0.06$, $\beta=4.02\pm0.32$  and $\sigma_0=200$\,km\,s$^{-1}$ \citep{tremaine02}. We adopt $\sigma_*\approx90$\,km\,s$^{-1}$ as representative of the bulge, a value in good agreement with previous optical measurements \citep{nelson95}, and obtain $M_{BH}=5.4^{+2.6}_{-1.8}\times10^6$\,M$_\odot$.

\subsection{Gas kinematics}
\label{gas_kin}

Although the gaseous velocity fields of Fig.~\ref{vel} present also a similar rotation pattern to that observed for  the stars, 
they are more disturbed, indicating that additional kinematic components -- not dominated by the gravitational potential of the bulge -- are present. Differences between the gas and stellar kinematics are also observed in the one-dimensional cuts along the major axis of the galaxy, shown in the top panels of Fig.~\ref{cut_kin}. To south-east of the nucleus the gaseous velocities are redshifted by $\approx$30~\kms\ relative to the stellar ones, while at $\approx$0\farcs5~north-west they are redshifted by $\approx$40~\kms. Some differences between the \feii, \h2\ and \pb\ kinematics are also observed, in particular between 1$^{\prime\prime}$ and 1\farcs5 north-west from the nucleus, where the \h2\ and \pb\ velocities are similar to the stellar and the \feii\ velocities are blueshifted by  $\approx$30~\kms.
A distinct kinematics for the \h2, \feii\ and \pb\ is also supported by the $\sigma$ maps (Fig.~\ref{sig}) and one-dimensional $\sigma$  cuts (top-right panel of Fig.~\ref{cut_kin}), which show that, at most locations, the \h2\ presents the smallest $\sigma$ values, followed by \pb, with  \feii\ presenting the highest $\sigma$ values.  
Differences between the kinematics of the ionized and molecular gas are also observed in 
the $h_3$ and $h_4$ maps (Figs.~\ref{h3} and \ref{h4}, respectively), as well as in the channel maps (Figs.~\ref{slice_feii}, \ref{slice_pb}, \ref{slice_h2}), which shows \feii\ emission at highest blue- and redshifts than \h2, indicating that the \h2 emission traces less disturbed gas than the \feii\ emission. 

The above conclusion is supported by other recent works. \citet{hicks09} have studied the \h2\ kinematics from the inner $\approx$100~pc of a sample  of 9 Seyfert galaxies using SINFONI IFU at the ESO Very Large Telescope (VLT) and concluded that it is dominated by rotation in a disc with typical radius of $\approx$30~pc and a comparable height. In previous studies by our group we have also found that while  the \feii\ emitting gas has important kinematic components attributed to gas extending to high galactic latitudes and in interaction  with a radio jet, the \h2\ kinematics is dominated by rotation in the galaxy disc \citep{rodriguez-ardila05a,riffel06,riffel08,riffel09,storchi-bergmann99,storchi-bergmann09}.

We can compare our results with previously published near-IR \citep{knop01} and optical \citep{bower95} gas kinematics  for Mrk\,1066 using long-slit spectroscopy. \citet{bower95} present  measurements of the centroid velocity and FWHM of the \ha, \hb, \oiii$\,\lambda5007\,\AA$, [O\,{\sc i}]$\,\lambda6300\,\AA$, \nii$\,\lambda6584\,\AA$, [S\,{\sc ii}]$\,\lambda6724\,\AA~$ emission lines along PA=134$^\circ$, obtained under a seeing of $\approx$0\farcs9.  All emission lines, with exception of the \oiii, present similar velocities which are consistent with our near-IR measurements. Nevertheless, our velocity field  reveals much more details, which we attribute to our better spatial resolution. The \oiii\ emission is blueshifted relative to  the other emission lines to north-west of the nucleus and redshifted to south-east. The presence of this kinematic component in our data is suggested by our  $h_3$ maps for \feii\ and \pb\ (bottom-left and top-right panels of Fig.~\ref{h3})  and one-dimensional cuts (bottom-left panel of Fig.~\ref{cut_kin}), which show positive values (red wings) to south-east of the nucleus  and negative values to north-west (blue wings). The presence of wings have also been observed by \citet{knop01} for the near-IR  emission lines along  PA=135$^\circ$ at a seeing of 0\farcs7. \citet{knop01} suggested in addition that these wings  are associated with the \oiii\ kinematic component of \citet{bower95}. In particular, at $\approx$1$^{\prime\prime}$ north-west of  the nucleus the \feii\ emission line is blueshifted by $\approx$30~\kms\  relative to the other near-IR emission lines, suggesting that the \feii\ emitting gas has the kinematical component observed in \oiii. 
In addition,  both present higher velocity dispersions when compared to other optical and near-IR emission lines with peak values observed at $\approx$1$^{\prime\prime}$  north-west of the nucleus, close to a radio hot spot. These results support the conclusion that at least part of the \feii\ emitting gas has the same kinematics of the \oiii\ emitting gas.

\begin{figure*}
\centering
\includegraphics[scale=1.0]{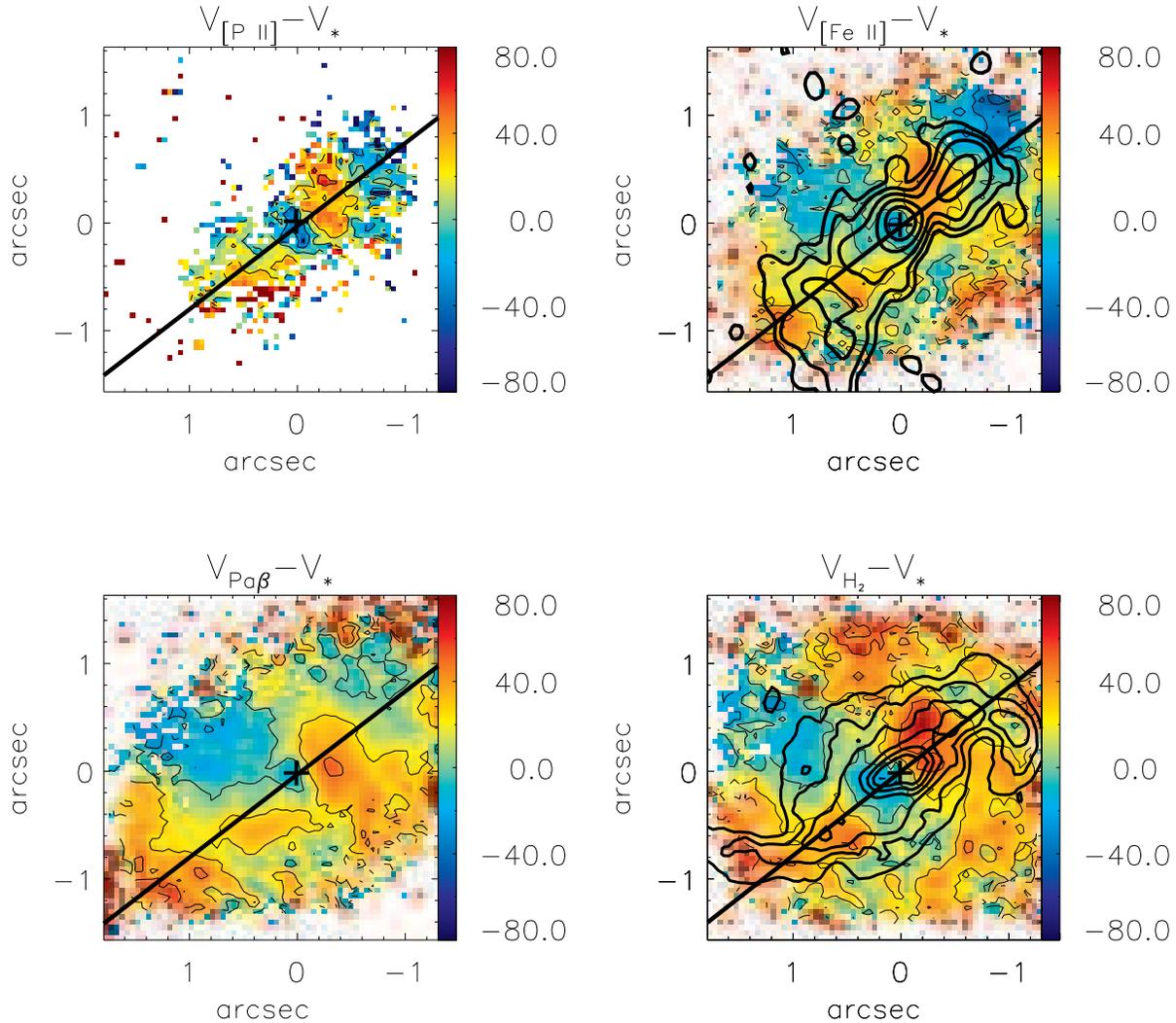}
\caption{Residual maps for the gas kinematics -- differences between the gaseous and stellar velocity fields for  \pii\ (top-left panel), \feii\ (top-right panel), \pb\ (bottom-left panel) and \h2\ (bottom-right panel). The thick contours overlaid to the \feii\ map are from the radio image, while those overlaid to the \h2\ map are from the HST optical continuum. The continuous line marks the line of nodes. 
}
\label{res}
\end{figure*}

As our goal here is to look for non-circular motions in the gas, we have subtracted the stellar velocity field (which is dominated by circular rotation) from the gaseous ones, considering also that we could not fit a circular model to the gas kinematics, which is too much disturbed. Figure~\ref{res} shows the resulting  maps for the the \pii\ (top-left panel), \feii\ (top-right), \pb\ (bottom-left) and \h2\ (bottom-right). Although these residual maps should be considered with caution, as the gas response to the gravitational potential differs from that of the stars \citep[e.g.][]{combes95}, we have used them only as guides, together with the gas centroid velocity fields and channel maps (Figs.~\ref{vel}, \ref{cut_kin}, \ref{slice_feii}, \ref{slice_pb} and \ref{slice_h2}) in order to look for non-circular velocity components. Considering all these maps, we tentatively identify the presence of two structures in the gas (besides the gas  in similar rotation to that of the stars): (i) a compact disc with excess velocity relative to that of the stars of $\approx$40~\kms\ within $\approx$0\farcs4 from the nucleus, more clearly observed in the \h2 residual map (bottom right panel of Fig.~\ref{res}); (ii) an outflow identified by excess blueshifts of $\approx$50~\kms\ at 1\farcs3 to north-west of the nucleus, close to the edge of the radio jet, more clearly observed in the \feii\ and \pii\ residual maps (top left and right panels of Fig.~\ref{res}). A similar structure in redshift is observed at $\approx$1$^{\prime\prime}$  south-east of the nucleus. We now discuss the observational signatures of these two components --  the compact disc and the outflow.

\subsubsection{The compact disc} \label{comp_i} 

This structure is observed not only in the residual map, but also directly in the H$_2$ velocity field (bottom-right panel of Fig.\,\ref{vel}), which shows an abrupt increase in velocity between the nucleus and $\approx\,0\farcs3$ ($\approx$\,70\,pc), while the stellar rotation shows a much shallower increase in the rotation velocity, what can also be seen in the one-dimensional cuts of Figure \ref{cut_kin}. The presence of this disc in \h2\ is further supported by the low velocity dispersions (see Figs.~\ref{sig} and \ref{cut_kin}), with values smaller than the stellar ones, observed closer to the nucleus in \h2\ than in the other emission lines.

Recent IFU observations of other Seyfert galaxies by other authors have also revealed the presence of circumnuclear \h2\ discs \citep{hicks09}. 
Following \citet{hicks09}, we estimate the dynamical mass within the compact disc radius (70~pc) from its kinematics as: 
\begin{equation}
 M_{dyn}\approx\frac{(V_{\rm rot}^2+3\sigma^2)\,r}{G},
\end{equation}
where $V_{\rm rot}=V_{\rm obs}/{\rm sin}(i)$ is the rotational velocity,  
 $V_{\rm obs}$ is the observed velocity, $G$ is the gravitational constant. From Fig.~\ref{cut_kin} we observe that at 0\farcs3 from the nucleus, the \h2\  $V_{\rm obs}\approx$65~\kms\ (average of the values observed to north-west and to south-east of the nucleus) and $\sigma \approx$~60\kms, resulting in $M_{dyn}\approx3.6\times10^8~{\rm M_\odot}$. Assuming a typical gas mass fraction of 10\% (see \citet{hicks09} for a discussion about the observed mass fraction), we estimate that the gas mass in the disc is $M_{gas}\approx3.6\times10^7~{\rm M_\odot}$. This mass is about one order of magnitude larger than those obtained by  \citet{hicks09} for a sample of 9 active galaxies, what can be due to the fact that the radius of the nuclear disc in Mrk\,1066 is at least two times larger than those in \citet{hicks09}. The mass derived above probably consists mostly of cold molecular gas since, as pointed out in \p1, the mass of hot molecular gas (which emits in the near-IR) is about 10$^4$ times smaller than the dynamical mass obtained here.  Radio observations indeed show the presence of large amounts of cold molecular gas with masses of 10$^{7}-$10$^{9}\,{\rm M_\odot}$ in the inner hundred parsecs of active galaxies \citep[e.g.][]{garcia-burillo05,boone07,krips07}. 

\vspace{0.1cm}
\noindent{ \it Feeding of the compact disc}
\vspace{0.1cm}


As pointed out in Section \ref{chamaps} the \h2\ channel maps  (Fig. \ref{slice_h2})  are not symmetric with respect  to the major axis in the blueshifted channels centred at $-104$\,\kms\ and $-42$\,\kms, where the flux distribution curves towards the north, and in the redshifted channels centred at 81\,\kms\ and 142\,\kms\ where there is a similar curvature  to the south (see the arrows in Figure~\ref{slice_h2}). In order to interpret these features, we now consider that the major axis of the galaxy  is at PA=128$^\circ$, running from the bottom left to the upper right in the panels of Fig. \ref{slice_h2}, as shown by the white line. Under the assumption that the spiral arms (outlined by the contours of the HST image superposed in Fig. \ref{stel}) are trailing, we conclude that the north-east is the far and the south-west is the near side of the galaxy. If the \h2\ emitting gas is in the galaxy plane, as suggested by its low velocity dispersion (bottom right panel of Fig.\,\ref{sig} and upper right panel in Fig.\ref{cut_kin}) we suggest that the curves in  the light distributions in the channels maps centred at $-104$\,\kms\ and $-42$\,\kms\ to the north-east, as well as in those centred at 81\,\kms\ and 142\,\kms\  to the south-west (see arrows) are due to inflows along spiral arms. We note that the same structures -- resembling spiral arms -- are observed also in the \h2\ velocity dispersion map (Fig.~\ref{sig}).

\begin{figure}
\centering
\includegraphics[scale=0.55]{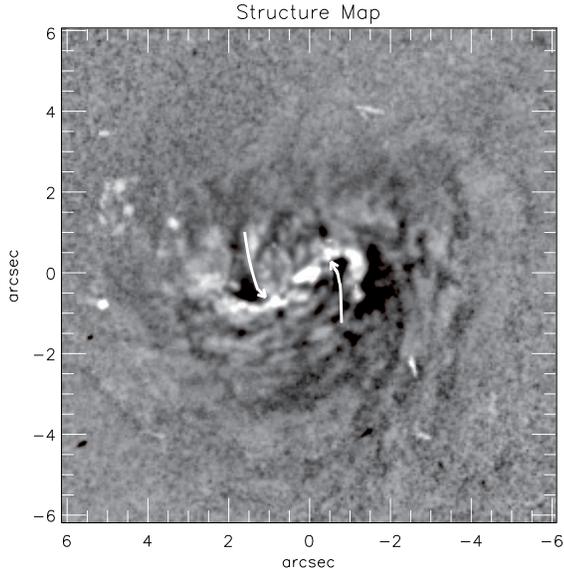}
\caption{Structure map of the inner region of Mrk\,1066 obtained from the HST image. The arrows  have been drawn at the locations they appear in Fig.\,\ref{slice_h2}, and approximately trace  nuclear spiral arms. 
}
\label{structure}
\end{figure}

Morphological nuclear spiral arms can also be seen in the optical HST F606W image of Mrk\,1066 \citep[see ][]{paper1}. A structure map, obtained from this image using the technique of \citet{pogge02} is shown in Fig.\,\ref{structure}. The arrows in this map approximately trace dark spiral arms and correspond to the same location we have drawn them in the \h2\ channel maps. In the structure map, the dark spiral arms are due to dust. It is expected that the \h2\ is associated to dusty regions, where it is protected from strong radiation which may destroy  the \h2\ molecule. This has been shown to be the case for NGC\,1097 by \citet{davies09}, who found velocity residuals in the \h2\ kinematics which were associated with dusty spirals \citep{prieto05}, interpreted as inflows towards the nucleus. Inflows were also observed along these spirals in ionized gas by \citet{fathi06}. \citet{lopes07}, using HST optical images for a matched sample of galaxies  with and without AGNs, found that, for the early-type subsample, all AGN hosts have circumnuclear structures of gas and dust, while this is observed in only 26\% of the  inactive galaxies. In many cases, the dusty structures observed by \citet{lopes07} resemble spiral arms, suggesting that they are the feeding channels of the AGN, as confirmed for a few cases in which the kinematics has been obtained. 

Inflows along nuclear spirals have been predicted in numerical simulations by \citet{maciejewski04a,maciejewski04b}, who demonstrated that, if a central SMBH is present, spiral  shocks can extend all the way to the vicinity of the SMBH and generate gas inflows consistent with the accretion rates inferred in local AGN. \citet{martini99} in their study of nuclear spirals in HST images have indeed concluded that they are not self-gravitating, being consistent with shocks in nuclear gas discs. Kinematical spiral arms
have also been observed in the central region of other Seyfert galaxies \citep{fathi06,storchi-bergmann07,riffel08,sanchez09,vandeVen09,davies09}. The modeling of the gaseous velocity field by a rotating disc plus spiral shocks shows that  the gas is slowly streaming towards the nucleus with a typical mass inflow rate of 10$^{-2}$--10$^1$\,M$_\odot$/yr \citep{davies09,vandeVen09,sanchez09}. Thus, the nuclear spiral arms seen in the \h2\ kinematics in Mrk\,1066 could also be the feeding channels of its SMBH, or at least of the compact disc discussed above. 

\subsubsection{The outflowing gas}\label{comp_iii}

The top-right panel of Fig.~\ref{res} shows blueshifts in the \feii\ residual map
of up to $-$80\,\kms\ at $\sim1^{\prime\prime}$ north-west of the nucleus, at the top edge of the radio jet. 
Similar residuals are observed for the \pii, and somewhat smaller for \pb\ and \h2\ (Fig.~\ref{res}). To the south-east the residual 
maps present similar redshifts in a region near the bottom edge of the radio jet.
We interpret these residuals as due to an outflow and in particular from regions   
of interaction between the radio jet and the ambient gas. This interpretation is supported by (i) 
the residuals being observed close to the ends of the radio jet, as mentioned above; (ii) the fact that the \feii\ emitting gas presents increased velocity dispersion values in regions  surrounding the radio structure. Such $\sigma$ enhancements are expected when gas is disturbed by a radio jet \citep[e.g.][]{dopita95,dopita96}; (iii) The $h_3$ map (Fig.~\ref{h3}) for the \feii\ presents its most negative values, of  $\approx -$0.3, in a region co-spatial with the blueshifts observed in the residual map; (iv) The velocity channel maps of  \feii\ are observed up to higher blueshifts and redshfits  than those for \pb\ and \h2. The flux distributions at these high velocities  are well correlated with the radio structure.  This behaviour is similar to that observed for the \oiii\ emission \citep{bower95}, interpreted as originated in outflowing gas in a bi-conical structure with axis approximately coincident with the orientation of radio jet, while the lower excitation optical line emission  is attributed to gas located in the plane of the galaxy.

The outflowing component is not more obvious in the near-IR because of the orientation of the outflow, which is close to the major axis of the galaxy, where the gas kinematics is dominated by rotation in the plane 
(as observed in the stellar velocity field of Fig.~\ref{stel}) and in the compact disc (Sec.\ref{comp_i}). The blueshifts described above represent a kinematic component which is superimposed on the galaxy rotation and on the compact disc component. The blueshifts at $\sim1^{\prime\prime}$ north-west  and redshifts at similar distances to the south-east can be interpreted as originating in a bi-conical outflow which shares the  \oiii\ kinematics but is outshined by emission from the galaxy disc and from the compact nuclear disc in the inner regions.

The bi-conical outflow is better observed in the \feii\ channel maps (Fig.~\ref{slice_feii}), which show the highest blueshifts (from $-300$ to $-$500\,\kms) to north-west of the nucleus and the highest redshifts observed to both sides of the nucleus. Comparing the channel maps with the \feii\ velocity field (Fig.~\ref{vel}) we conclude that the north-west  blueshifts and south-east redshifts are due to outflows, as the galaxy rotation is in the opposite direction, showing blueshifts to the south-east and redshifts to the north-west. At intermediate velocities (80--220~\kms) both  redshifts and blueshifts are observed to both sides of the nucleus confirming the presence of two velocity components: rotation and outflows. A similar behaviour is observed for the \pb\ channel maps.  Lower velocity channels are  consistent with emission from gas in the plane of the galaxy. The \h2\ channel maps are dominated by these lower velocities, supporting the interpretation that the \h2\  is located in the galaxy plane. 

Under the assumption that part of the \feii\ emitting gas shares the \oiii\ kinematics, we can estimate the opening angle of the cone directly from Fig.~\ref{res} as being $\alpha\approx20^\circ$. As the radio jet is oriented almost along the galaxy major axis, we can constrain the angle of the bi-cone axis relative to the plane of the sky as being close but somewhat larger than zero ($\theta\gtrsim0^\circ$), since the north-west side of the bi-cone must be oriented towards us (although apparently by a small angle) in order to produce excess blueshifts there while the south-east side of the bi-cone is directed away from us in order to produce the observed redshifts. 

 We now estimate the mass outflow rate across a circular cross section along the bicone located at 1\farcs3 from the nucleus, which has a radius $r\approx$0\farcs23\,$\approx54\,$pc for a bi-cone opening angle of 20$^\circ$,  
corresponding to an area of  $A\approx8.7\times10^{40}$~cm$^{2}$. The component of the velocity of the outflowing gas along the bi-cone axis ($v_{\rm out}$) 
is related to the observed velocity ($v_{\rm obs}$) by $v_{\rm out}=v_{\rm obs}/{\rm sin}\,\theta$, where 
$\theta$ is the angle between the bi-cone axis and the plane of the sky, and thus the mass outflow rate can be obtained from

\begin{equation}
 \dot{M}_{\rm out} = \frac{2 m_p N_e v_{\rm obs} f A}{sin\,\theta},
\end{equation}
where the factor 2 is included in order to account for the outflows to both sides of the nucleus. Assuming $N_e=500\,{\rm cm^{-3}}$, $f=0.01$ and  $v_{\rm out}=50$~\kms, directly form Fig.~\ref{res}, we obtain $\dot{M}_{\rm out}\approx1\times10^{-2}/{\rm sin\,\theta}~{\rm M_\odot\, yr^{-1}}$. 

In order to estimate the value of $\theta$, we consider the observation that the \oiii\ image presents a collimated structure to north-west of the nucleus and a fainter emission to the south-east \citep{bower95},  suggesting that the bi-conical outflow makes a small angle with the plane of the sky, with the south-east side apparently obscured by the galaxy plane. A large angle would produce a less collimated emission.

In order to quantify the mass outflow rate, we assume $\theta=10^\circ$. In this case the adopted geometry implies that the far wall of the north-west cone  and the near wall of the south-east  cone are exactly in the plane of the sky. With this value of $\theta$ we obtain  $\dot{M}_{\rm out}\approx6\times10^{-2}~{\rm M_\odot\, yr^{-1}}$.

The mass outflow rate obtained for Mrk\,1066 can be compared with those obtained for other active galaxies. 
\citet{veilleux05} report values of $\dot{M}_{\rm out} \approx0.1-10~{\rm M_\odot\, yr^{-1}}$ for warm ionized gas outflow from luminous active galaxies, 
while \citet{barbosa09} report $\dot{M}_{\rm out} \approx(1-50)\times10^{-3}~{\rm M_\odot\, yr^{-1}}$ from optical IFU observation of a sample of six Seyfert 
galaxies assuming and electron density of $N_e=100~{\rm cm^{-3}}$. \citet{storchi-bergmann09b} found $\dot{M}_{\rm out} \approx2~{\rm M_\odot\, yr^{-1}}$ 
from the nucleus of NGC\,4151 using similar NIFS observations of its circumnuclear region, while \citet{crenshaw07} obtained an outflow rate $\approx$\,6 
times smaller using blueshifted absorption lines in the UV spectrum of NGC\,4151. In \citet{riffel09}
 we estimated $\dot{M}_{\rm out} \approx5\times10^{-2}~{\rm M_\odot\, yr^{-1}}$ for the Seyfert~2 nucleus of NGC\,7582 
for the ionized gas using K-band IFU spectroscopy. The mass outflow rate estimated for Mrk\,1066 is thus within 
the range of observed values for active galaxies, being 2 times smaller than the lowest ones obtained by \citet{veilleux05}, $\approx$\,30 times smaller than the one obtained
 for NGC\,4151 \citep{storchi-bergmann09b,crenshaw07}, about 10 times larger than the values obtained by \citet{barbosa09} and similar to the outflow rate of NGC\,7582 \citep{riffel09}.

\subsubsection{Feeding {\it vs} feedback in Mrk\,1066}

\begin{figure*}
\includegraphics[scale=1.0]{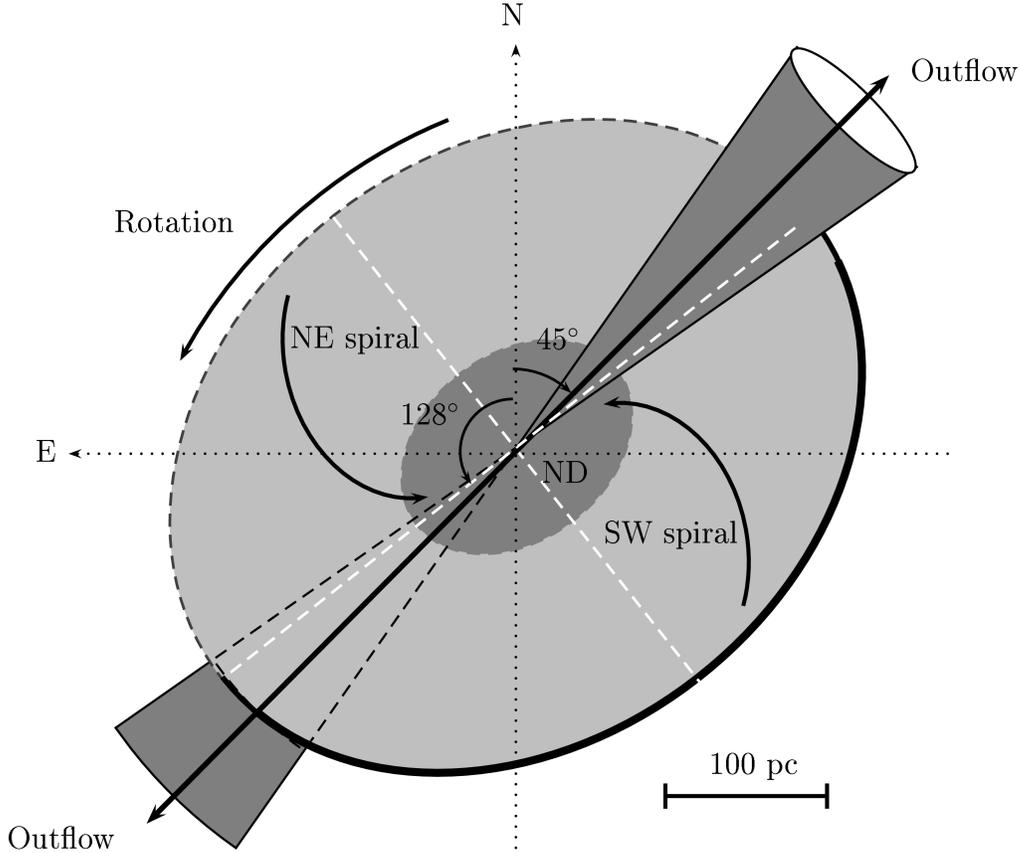}
\caption{Physical scenario for the circumnuclear region of Mrk\,1066 \citep{knop01}. Most of the near-IR line emission originates in a rotating disc with major axis coincident with that of the stellar disc (white dashed line along PA=128$^\circ$). 
The \oiii\ line emission originates mostly from gas in the bi-conical outflow with redshifts to south-east and blueshifts to north-west of the nucleus, 
opening angle $\approx20^\circ$ and oriented along the PA of the radio jet (arrows along PA$\approx$135/315$^\circ$). The outflowing component observed in the near-IR emission lines also originates in the bi-conical outflow. The arrows shown along the minor axis represent the nuclear spiral arms observed in Fig.~\ref{res}.}
\label{scenario}
\end{figure*}

\citet{knop01} -- using long-slit spectroscopy -- proposed that most of the near-IR line emission in Mrk\,1066, 
as well as the optical lines of \nii, \sii, \oi\ and H\,{\sc i}, originate in a rotating disc in the plane 
of the galaxy with a major axis oriented along  PA$\approx120^\circ$, while the wings of the near-IR 
emission lines and the optical \oiii\ emission originate from the bi-conical outflow, as illustrated in their Fig.~8.

Our observations have 2D coverage, as well as higher spectral and spatial resolutions than those of \citet{knop01} and can be used to better constrain the orientation of the cone and major axis of the disc and we can thus further constrain the scenario proposed by \citet{knop01}. In Fig.~\ref{scenario} we present a cartoon illustrating our proposed physical scenario for the circumnuclear region of Mrk\,1066. This cartoon is inspired on Fig.~8 of \citet{knop01}, in which the high ionization emission lines -- such as the optical \oiii\ lines -- originate in outflowing gas along the bi-cone oriented along the radio axis (PA$\approx135/315^\circ$) and with an opening angle of $\approx20^\circ$.  

In agreement with \citet{knop01}, we support that the low-excitation line emission (from \h2, H\,{\sc i}, \feii, \pii, \nii, \oi, \sii) is dominated by gas rotating in the galaxy disc, with similar kinematics to that of the stars: the line of nodes is oriented along $PA\approx128^\circ$ and the disc inclination is $i\approx38^\circ$  relative to the plane of the sky.
 But we can also identify in the near-IR emission lines a kinematic component originating in the bi-conical outflow (which 
dominates the \oiii\ emission), as evidenced by the blueshifts to north-west and redshifts to south-east of the nucleus in Fig.~\ref{res}. The contribution of this component to the gas kinematics increases in importance from \h2$\rightarrow$\pb$\rightarrow$\feii. This can also be observed in the channel maps: while the presence of outflows combined with rotation is clear in the \feii\ and \pb\ channel maps, the \h2 channel maps seem to show only rotation. These kinematics suggest that part of the ionized gas extends to high galactic latitudes, where it contributes to the bi-conical outflow, while the molecular gas is mostly restricted to the galaxy plane.

This picture -- in which the \h2 originates mostly from the galaxy plane, while the ionized gas is observed also in outflows extending to high galactic latitudes --  is in good agreement with results we have obtained for other Seyfert galaxies \citep[e.g.][]{riffel06,riffel08,riffel09,storchi-bergmann99,storchi-bergmann09,storchi-bergmann09b}.

Another interesting new result is the observation of a compact nuclear disc, with radius $\approx$70\,pc,
which is rotating faster than the underlying stellar disc. We interpret this nuclear disc as a gaseous structure which will soon give origin to the formation of new stars.  We are probably witnessing the formation of a nuclear stellar disc, such as those observed as low stellar velocity dispersion structures surrounding active galactic nuclei \citep{davies07} or as high surface brightness structures in previous HST imaging studies \citep[e.g.][]{lopes07,pizzella02}. This compact disc is apparently being fed by molecular gas flowing along nuclear spirals, as evidenced by the blueshifts observed at the far side of the galaxy and redshifts in the near side (Fig.~\ref{slice_h2}) 
similarly to what we have observed in the Seyfert galaxy NGC\,4051 \citep{riffel08}. 


Finally, we now compare the outflow mass rate, obtained in Sec.\ref{comp_iii}, with the accretion rate necessary to power the AGN at the nucleus of Mrk\,1066.

The mass accretion rate to feed the active nucleus can be obtained from
\begin{equation}
 \dot{m}=\frac{L_{\rm bol}}{c^2\eta},
\end{equation}
where $L_{\rm bol}$ is the nuclear bolometric luminosity, $\eta$ is the efficiency 
of conversion of the rest mass energy of the accreted material into radiation and $c$ 
is the light speed. The bolometric luminosity can be approximated by 
$L_{\rm bol} \approx 100L_{\rm H\alpha}$, where $L_{\rm H\alpha}$ is the \ha\ nuclear luminosity \citep[e.g.][]{ho99,ho01}. The \br\ nuclear flux measured for an circular aperture with 0\farcs25 radius is $F_{\rm Br\gamma}\approx4.2\times10^{-15}\,{\rm erg\, s^{-1} cm^{-2}}$. In \p1, we obtained a reddening of $E(B-V)\approx1$ for the nucleus of Mrk\,1066 using the \br/\pb\ line ratio. Correcting the \br\ flux for this reddening using the law of \citet{cardelli89}, we obtain  $F_{\rm Br\gamma}\approx5.8\times10^{-15}\,{\rm erg\, s^{-1} cm^{-2}}$. Assuming a temperature $T=10^4$~K and an electron density $n_e=10^2~{\rm cm^{-3}}$, the ratio between  \ha\ and \br\ is predicted to be $F_{\rm H\alpha}/F_{\rm Br\gamma}=103$ \citep{osterbrock06}. Thus $L_{\rm H\alpha}\approx1.7\times10^{41}\,{\rm erg\, s^{-1}}$ for a distance to Mrk\,1066 of $d=48.6$~Mpc and the nuclear bolometric luminosity is estimated to be $L_{\rm bol}\approx1.7\times10^{43}\,{\rm erg\, s^{-1}}$. Assuming $\eta\approx0.1$, which is a typical value for a ``standard'' geometrically thin, optically thick accretion disc \citep[e.g.][]{frank02}, we obtain a mass inflow rate of $\dot{m}\approx3\times10^{-3}~{\rm M_\odot\, yr^{-1}}$.


The mass outflow rate ($\dot{M}_{\rm out}$) in the NLR is about one order of magnitude larger than $\dot{m}$, a ratio comparable to those observed for other Seyfert galaxies \citep{barbosa09,riffel09,storchi-bergmann09b}, which indicates that most of the outflowing gas in the NLR of active galaxies does not originate in the AGN but in the surrounding interstellar medium from the galaxy plane, which is pushed away by the nuclear outflow. 

\section{Summary and Conclusions}

We used integral field J and K$_{\rm l}$ bands spectroscopy from the inner $\approx$350\,pc radius of the Seyfert galaxy Mrk\,1066, obtained with the Gemini NIFS at a spatial resolution of $\sim$35\,pc and spectral resolution of $\sim$40\,\kms, to map the gaseous and stellar kinematics. Our main conclusions are:

\begin{itemize}

\item The stellar velocity field is dominated by circular rotation in the plane of the galaxy, but shows an S-shape distortion along the galaxy minor axis which we attribute to the presence of an oval structure in the galactic potential.

\item The oval structure is also delineated by a partial ring of low velocity dispersion ($\sigma_*\approx$50\,\kms)  at $\approx$\,230\,pc from the nucleus, which is due to intermediate-age stars immersed  in the old bulge, as concluded from stellar population synthesis \citep{paper2}.

\item The bulge velocity dispersion is $\sigma_*\approx90$\,\kms, implying in a black hole mass of $M_{\rm BH} \approx 5.4 \times 10^6$\,M$_\odot$.   

\item The gaseous kinematics shows circular rotation similar to that of the stars with two additional kinematic components: (i) a compact nuclear disc, more clearly observed in \h2\ emission, with radius $r=70\,$pc and gas mass of $\sim10^7\,{\rm M_\odot}$; (ii) outflows associated with the radio jet, presenting a mass outflow rate of $\approx6\times10^{-2}{\rm M_\odot yr^{-1}}$, more clearly observed in \feii\ emission.

\item The compact  \h2\ nuclear disc presents the smallest gas $\sigma$ values ($\le$\,70\,\kms), consistent with an origin in the galaxy plane and is apparently being fed via nuclear spiral arms, observed in molecular gas at similar velocity dispersion to that of the compact disc. This disc may not only be the feeding source of the AGN but probably will be consumed in star formation, leading to the formation of a nuclear stellar disc, a structure frequently observed in galaxy bulges.

\item The \feii\ emission presents the largest $\sigma$ values (up to 150\,\kms) and the highest blue- and redshifts, of up to 500\,\kms, while the highest stellar rotation velocity is only $\approx$\,130\,\kms. The high velocities and velocity dispersions are found at the tip and around the radio jet, which is oriented approximately along the galaxy major axis, with the north-west side slightly tilted towards us. The outflow rate is about one order of magnitude higher than the accretion rate, indicating that most of the outflow consists of entrained gas from the galaxy plane by a nuclear outflow.

\end{itemize}

\section*{Acknowledgments}
We thank the referee for valuable suggestions which helped to improve the present paper, as well as Dr. Neil M. Nagar for sending the radio continuum image.
 This work is based on observations obtained at the Gemini Observatory, 
which is operated by the Association of Universities for Research in Astronomy, Inc., under a cooperative agreement with the 
NSF on behalf of the Gemini partnership: the National Science Foundation (United States), the Science and Technology 
Facilities Council (United Kingdom), the National Research Council (Canada), CONICYT (Chile), the Australian Research 
Council (Australia), Minist\'erio da Ci\^encia e Tecnologia (Brazil) and south-eastCYT (Argentina).  
This research has made use of the NASA/IPAC Extragalactic Database (NED) which is operated by the Jet
 Propulsion Laboratory, California Institute of  Technology, under contract with the National Aeronautics and Space Administration.
This work has been partially supported by the Brazilian institution CNPq.

\label{lastpage}

\end{document}